\documentclass[conference]{IEEEtran}
\IEEEoverridecommandlockouts
\usepackage{amsmath,amssymb,amsfonts}
\usepackage{algorithmic}
\usepackage{subfigure}
\usepackage{graphicx}
\usepackage{multirow}
\usepackage{adjustbox}
\usepackage{url}
\usepackage{textcomp}
\usepackage[
backend=biber,
style=numeric,
sorting=nyt,
]{biblatex}
\usepackage{xcolor}
\def\BibTeX{{\rm B\kern-.05em{\sc i\kern-.025em b}\kern-.08em
    T\kern-.1667em\lower.7ex\hbox{E}\kern-.125emX}}
    
\newcommand{\nop}[1]{}
\begin{document}

\title{%Nonlinear Models Analyzing Transaction Handling in Bitcoin
An Analysis of Transaction Handling in Bitcoin
%Transaction Handling in Bitcoin: Is It Predicable?
}
%\author{}
\author{\IEEEauthorblockN{Befekadu G. Gebraselase, Bjarne E. Helvik, Yuming Jiang}
\IEEEauthorblockA{\textit{Department of Information Security and Communication Technology} \\
\textit{NTNU, Norwegian University of Science and Technology, Trondheim, Norway} \\
\{befekadu.gebraselase, bjarne, yuming.jiang \}@ntnu.no}

}

\maketitle

\begin{abstract}

 Bitcoin has become the leading cryptocurrency system, but the limit on its transaction processing capacity has resulted in increased transaction fee and delayed transaction confirmation. As such, it is pertinent to understand and probably predict how transactions are handled by Bitcoin such that a user may adapt the transaction requests and a miner may adjust the block generation strategy and/or the mining pool to join. To this aim, the present paper introduces results from an analysis of transaction handling in Bitcoin. 

Specifically, the analysis consists of two-part. The first part is an exploratory data analysis revealing key characteristics in Bitcoin transaction handling. The second part is a predictability analysis intended to provide insights on transaction handling such as (i) transaction confirmation time, (ii) block attributes, and (iii) who has created the block. The result shows that some models do reasonably well for (ii), but surprisingly not for (i) or (iii).

\end{abstract}

\begin{IEEEkeywords}
 Bitcoin, Transaction handling,  Linear and nonlinear prediction models, Classification, Machine Learning, Artificial Intelligence %Throughput
\end{IEEEkeywords}

\section{Introduction}

Bitcoin is the first and the largest decentralized electronic cryptocurrency system that uses blockchain technology \cite{Nakamoto}. It adapts a cryptographic proof of work (PoW) mechanism that allows anonymous peers to create and validate transactions through the underlying peer-to-peer (P2P) network.  The peers that maintain and update the chain of blocks are called miners \cite{miningPool, miningview}. In addition to transaction generation by user nodes, transaction handling in Bitcoin is done by the full nodes,  among which, the miners play a central role: They find the mathematical puzzle to generate a valid block confirming the related transactions. % and hence are the backbone of the overall system. 

Due to the design and structure of  proof of work (PoW) in Bitcoin, the difficulty of finding the mathematical puzzle increases exponentially, every 2016 blocks. As a consequence,  independent miners struggle to find the puzzle. This has forced miners to collaborate to form a team to find the puzzle through a combined computational effort, a {\em mining pool} \cite{miningEvolu}, and earn a reward, depending on their overall mining power share and the reward mechanism and policy of the mining pool \cite{strategy} \cite{Profit}. 
The mining pools' behavior significantly affects the Bitcoin end users since the mining pools process most of the users' transactions: The throughput of Bitcoin depends partially on those major miners \cite{miningview}. Additionally, as the number of users increases, the system's internal traffic of transaction handling escalates faster than expected, and at the same time, the throughput requirement increases proportionally with the number of users. 

This paper investigates how transactions are handled by the Bitcoin system. %with a particular focus on if, or to what extend, this transaction handling may be predicted and the corresponding mining pools distinguished. 
The aim is to, through analyzing transaction handling, provide valuable insights to both users and miners: 
\begin{itemize}
    \item A user may expect when his/her transaction will be confirmed and hence choose an appropriate time to request a transaction to reduce the waiting time.
    \item A miner may define block generation strategies that utilize the current state of the system.
    \item A miner may also explore which mining pools are more recognizable in the block generation and use this knowledge to join or dis-join a mining pool. 
\end{itemize}
Specifically, through an exploratory data analysis, we reveal key transaction handling characteristics and provide answers to several fundamental transaction handling questions, such as, what is the current throughput, how frequently blocks are generated, how long it takes for a transaction to be approved, and who has created a block. Besides, through a predictability analysis on throughput related features and classification of mining pools, we provide additional insights on these fundamental questions. 

%Specifically, in addition to reporting key transaction handling related characteristics, we conduct a prediction study on throughput related features and apply decision tree based classification to distinguish mining pools. Through these, we provide answers / insights on several fundamental, transaction handling aspects and particularly if they are predicable: what is the current throughput, how frequently blocks are generated, how long it takes for a transaction to be approved, and who has created the block.

The investigation is based on a dataset collected at a Bitcoin full node which contains transaction handling information over a period of 543 days from 7th March, 2019 to 31st August 2020. As a highlight, the dataset includes locally available information that cannot be found on the public ledger blockchain. %related to 80408 blocks and 160 million transactions 
%This paper's contributions include {\bf (i)} a statistical study of key transaction handling related characteristics where  the contribution of main miners / mining pools is particularly factored in; {\bf (ii)} a history-based prediction study of Bitcoin's throughput related features such as block size and number of transactions in a block; {\bf (iii)} a history-based classification study of mining pools that also takes into account the potential impact of time. 
The results indicate that with a proper prediction model taking into account both internal and external factors, the prediction performance can be appealing for block size and number of transactions in a block, as well as for block generation intensity. However, in terms of predicting when a next block will be generated and a transaction be approved, the effort does not lead to conclusive observation. In addition, also surprisingly, in predicting / classifying the mining pool, clear distinguishing is only found for one specific mining pool, the F2Pool. Discussion is provided for these findings, including the surprising ones, with the help of findings from the exploratory analysis.

%The time between consecutive block generation follows an exponential distribution \cite{TransBitcoin, BitcoinBlockchainDynamics, trasactionConfirmation}. Predicting the next block generation time is difficult. To address this problem, in this work, we consider the block inter-generation time as equidistant.  To this extent, we defined throughput as the number of transactions processed per second.   

%The rest of the paper is organized as follows. Section \ref{work} illustrates how the Bitcoin process flow works. Then, Section~\ref{sec-dat} presents the dataset  collection and feature set analysis. Following that, the paper's problem definition and modeling are considered in Section \ref{sec-model}. Next, Section~\ref{sec-met} discusses the proposed methodology, highlighting the adopted statistical and artificial intelligence techniques. After that, results are discussed in Section~\ref{sec-res}. The current state of the art is covered in section~\ref{sec-stateArt}. Finally, Section~\ref{sec-con} concludes the paper and outlines the research directions for future extensions. 

The rest of the paper is organized as follows. Section \ref{work} illustrates the workflow of transaction handling in Bitcoin, and introduces the dataset used in the analysis. 
Then, Section~\ref{sec-met} introduces our analysis approach, highlighting the adopted statistical and artificial intelligence techniques.
Following that, an exploratory analysis on the dataset is conducted and results are reported and discussed in Section~\ref{sec-sa}. Next, Section~\ref{sec-pa} reports results and findings from the predictability study. 
The current state of the art is covered in Section~\ref{sec-stateArt}. Finally, Section~\ref{sec-con} concludes the paper. 
 
 \section{Bitcoin Transaction Handling: Workflow and Dataset}%{Bitcoin Transaction Handling: Workflow, Dataset and Analysis Approaches} %Prediction Models
 \label{work}
 \subsection{Workflow}
Bitcoin is a distributed ledger platform that enables information about transactions to be distributed than centralized, where the ledger is the Bitcoin blockchain that records the transactions. In Bitcoin, all full nodes, also called miners, take part in creating and validating/invalidating transaction blocks and propagating such information, independently~\cite{Nakamoto}. Specifically, the users generate transactions for being processed, and the distributed ledger components, i.e. the full nodes or miners, work together to generate and validate transaction blocks and add them to the blockchain.

\begin{figure}[htb!]
\centering
  \includegraphics[width=\linewidth,height=0.6\linewidth]{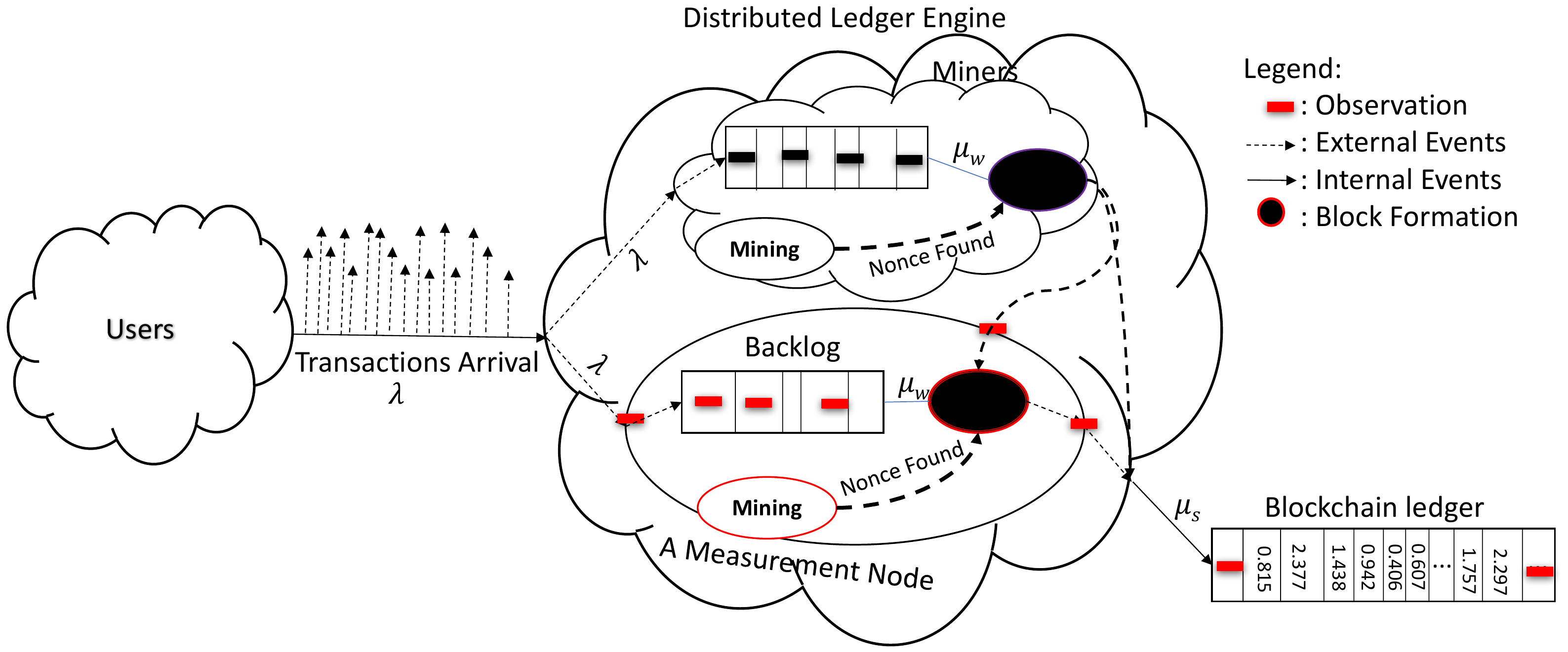}
  \caption{An illustration of the work flow of Bitcoin}%  
  \label{intro}
\end{figure}

Fig.\ref{intro} illustrates the workflow of transaction handling in Bitcoin, which includes transaction arrival, block formation, propagation and validation. Briefly, after transactions are generated by the users, they are sent to all full nodes for validation. At a full node, upon the arrival of a transaction, the node stores the transaction in its mining pool, called mempool in Bitcoin, waiting for confirmation. 

In addition, a full node may choose unconfirmed transactions in the backlog to pack into a new transaction block, and perform mining to find the mathematical puzzle given by the Bitcoin to gain the right to add the block to the ledger. 
If the puzzle finding is successful, this newly generated block is added to the blockchain, and this information is sent to all the nodes.  %

At each node, the validity of the newly generated block is checked. If the validity is confirmed with consensus, the updated blockchain is accepted and the transactions in the new block are validated. Such validated transactions are removed from the mempool at each full node that then repeats the above process. Note that, while the above description is brief, the essence of the workflow is kept. For more details about how Bitcoin works, the original introduction ~\cite{Nakamoto} is the best source.

\subsection{Dataset}  \label{sec-dat} % and Data Analysis
%This section covers two parts. The first part illustrated how the dataset was collected from the setup.  The second subsection makes some data analysis of the collected dataset feature set, further exploring the block attributes/features relative to the block generation time. 
%\subsection{Datasets}
To analyze transaction handling in Bitcoin, we implemented a server installation of a full Bitcoin node to collect related information. The information has two parts. One part  records information from the ledger that is globally available, called the {\em global information part}. Another part records locally available information about the backlog status of the mempool. This part is called the {\em local information part}.

More specifically, the global information part includes, for each block $i$ on the blockchain, the number of transactions ($n_i$) in the block, its miner ($m_i$), the size of the block in bytes ($s_i$), the timestamp or generation time of the block ($T_i$), and the average per-transaction fee of the block ($f_i$). The local information part records the mempool' status ($ms_i$) in terms of size and fee of backlogged transactions in mempool when each block $i$ is received at our full node. %A brief summary of these focused features is also shown in Fig.~\ref{deployment}. 
%This block contains attributes including the number of transactions in a block ($n_i$), the size of the block in bytes ($s_i$), the timestamp of the block ($T_i$), the average fee per block ($f_i$), and miner ($m_i$). 

In total, the dataset consists of information related to 80,408 Bitcoin blocks with more than two hundred million (203432240) transactions for a period of 543 days from 7th March 2019 to 31st August 2020.

\section{The Analysis Approach}\label{sec-met}
%Methodology{Prediction Models}

The dataset is essentially a composition of time series. We hence employ time series analysis on the dataset to provide insights and/or gain findings about transaction handling in Bitcoin. In the rest, the following time series are specifically used: 
$y=[y_1,y_2,\ \ldots., y_M]$, $x=[x_1,x_2,\ \ldots., x_M]$, $c=[c_1,c_2,\ \ldots., c_M]$, and $D= \{\{y_1, x_1, c_1\}, \{y_2, x_2, c_2\}, \ldots, \{y_M, x_M, c_M\}\}$, 
%\begin{eqnarray}
%y &=& [y_1,y_2,\ \ldots., y_M] \nonumber\\ x &=& [x_1,x_2,\ \ldots., x_M] \nonumber \\ 
%c &=& [c_1,c_2,\ \ldots., c_M] \nonumber \\
%D &=& \{\{y_1, x_1, c_1\}, \{y_2, x_2, c_2\}, \ldots, \{y_M, x_M, c_M\}\} \nonumber
%\end{eqnarray}
defined with: 
\begin{equation}
    \begin{split}
y_i= \{ s_i, n_i\} \\
x_i=\{Td_i, f_i, ms_i\}\\
c_i= \{ m_i\} %\\
%D= \{\{y_1, x_1, c_1\}, \{y_2, x_2, c_2\}, \ldots, \{y_M, x_M, c_M\}\}\; \\ M=80408 \;
    \end{split}
\end{equation}
where $Td_i \equiv T_i - T_{i-1}$ denotes the inter-block time, $s_i, n_i, f_i, ms_i$ and $m_i$ are defined in the previous section, and $M=80408$ representing the total number of blocks in the dataset.

Our analysis consists of two parts. In the first part, i.e., Section~\ref{sec-sa}, the focus is on revealing fundamental characteristics and/or basic statistical properties of transaction handling related time series, using exploratory data analysis techniques such as histogram, scatter plot and curve fitting. 

In the second part of the analysis, i.e. Section~\ref{sec-pa}, the focus is on investigating if / how Bitcoin transaction handling may be predicted. To this aim, both classical and modern time series forecasting approaches are considered for prediction of various transaction related attributes. In addition, a decision tree based classification approach is adopted for miner inference. The following subsections give an introduction of these approaches.

\subsection{Autoregressive models for forecasting} %ARIMA Models
For time series forecasting, a large number of approaches are available, including both classical ones and modern artificial intelligence (AI) based approaches \cite{faloutsos2019classical}. 

For the former, we tested various autoregressive (AR) models. Due to their generally better performance, this paper focuses on   
ARIMA (AutoRegressive Integrated Moving Average) and ARIMAX (Autoregressive Integrated Moving Average with Exogenous input). Equations (\ref{ARIMA}) and (\ref{ARIMAx}) define these models respectively, where $B$ is the backshift operator and $\nabla$ the difference operator. %(YM comment: Is $T_i=t$ in $({y}^+_{i}|T_i=t)$ and later (5) and (6) really needed for the explanation of (\ref{ARIMAx}) or (5) / (6)?)
\begin{align} \label{ARIMA}
\begin{split}
     {y}^+_{i} = \phi_{1}{y}_{t-1} + \cdots + \phi_{p}{y}_{i-p} 
     + \theta_{1}\varepsilon_{i-1} +  \cdots + \\ \theta_{q}\varepsilon_{i-q} +  \varepsilon_{i}, \\
     \Phi(B)\nabla^{d}{y}^+_{i}=\Theta(B)\varepsilon_{i},
\end{split}
\end{align}
\begin{align} \label{ARIMAx}
\begin{split}
      ({y}^+_{i}|T_i=t)=  \phi_{1}\{{x}_{i-1}, {y}_{i-1}\} + \cdots + \phi_{p}\{{x}_{i-p}, {y}_{i-p}\} \\
     + \theta_{1}\varepsilon_{i-1, t_{i-1}} + \cdots + \theta_{q}\varepsilon_{i-q, t_{i-q}} + \varepsilon_{i, t_i}, \\
     \Phi(B)\nabla^{d}({y}^+_{i}|T_i=t)=\beta x_i + \Theta(B)\varepsilon_{i,t_i},
\end{split}
\end{align}
where $({y}^+_{i}|T_i=t)$ or $ {y}^+_{i}$) is the predicted block, $E(\varepsilon_{i,t_i})=0$, Var($\varepsilon_{i,t_i})$ = $\sigma^2,$ $ \nabla^d$=(1-B)$^d$ is difference factor,  $\nabla ^{d}({y}^+_{i}|T_i=t)$ is the sequence of $y_i$ by $d$ times differed,  $\Phi(B)$= 1$-\phi_1B, \dots, \phi_pB^p$ is an auto regressive coefficient polynomial, and $\Theta(B)$=1$-\theta_1B, \dots, \theta_qB^q$ is a moving smoothing coefficient polynomial of the smooth invertible autoregressive moving average model ARMA $(p, q)$.

To assess the forecasting performance, we use mean average error (MAE) and root mean square error (RMSE), which are respectively defined as: with $e_i =  y_i-y_i^+$, 
%The difference between the measured and forecasted block values is the forecast error $e(i)$ or $(e(i, t_i$)). Equation (4) shows the forecast error considering both with and without external factors and the performance metrics, Mean Square Error (MSE) and Mean average error (MAE). 
\begin{equation}
\begin{split}
%      e_i =  y_i-y_i^+ \\
%      e(i, t_i) =  (y_i, t_i)-(y_i|T_i=t_i)^+ \\
      MAE=\frac{\sum_{i=1}^{N} |e_i|}{N} \\
%      MSE=\frac{\sum_{i=1}^{N} e_r^2}{N} \\
      RMSE=\sqrt{\frac{\sum_{i=1}^{N} e_i^2}{N}}%\sqrt{MSE}
\end{split}
\end{equation}
where 
%$y_i$ denotes the actual or measured value and $y_i^+$ denotes the value from prediction for the $i$-th data point in the time series $y$, and 
$N$ denotes the number of predicted data points. 

\subsection{AI-based forecasting models}

For AI-based models, NAR (nonlinear autoregressive neural network) and NARX (nonlinear Autoregressive Network with Exogenous Inputs) are chosen because they have a feedback connection that encloses several layers of the network, which uses memory to remember the time series's past values to get better performance~\cite{Nonlinear}~\cite{NarxModel}. Additionally, the models have nonlinear filtering that helps to capture the dynamic fluctuations of the input values. 

Equations (\ref{nareq}) and (\ref{narxeq}) describe NAR and NARX network's function to predict a particular value of data series $y^+_i$ using $p$ previous values of $y$ and $x$.
\begin{align} \label{nareq}
(y^+_{i})=f_{\rm{NAR}}(y_{i-1},y_{i-2}, \ldots.,y_{i-p})
\end{align}
\begin{equation}\label{narxeq}
    \begin{split}
        (y^+_{i}|T_i=t)=f_{\rm{NARX}}(\{x_{i-1},y_{i-1}\},\{x_{i-2},y_{i-2}\}, \\
\ldots.,\{x_{i-p},y_{i-p}\}) 
    \end{split}
\end{equation} 
The functions $f_{\rm{NARX}}$ and $f_{\rm{NAR}}$ in (\ref{nareq}) and (\ref{narxeq}) are unknown, and the neural network training approximates the function by optimizing the network weights and neuron bias. %, where $y_i$ predicts series for given $p$ past values of $y_i$.  
The NAR and NARX model uses Levenberg-Marquardt, Bayesian regularization, and scaled conjugate gradient training algorithms to train the model \cite{neuralNStateofart}. Specifically, Bayesian regularization (BR) is used to conduct the analysis. BR minimizes a combination of squared errors and weights then determines the correct combination to produce a network that generalizes well. It uses network training function Levenberg-Marquardt to optimize network weights and neuron bias. The Levenberg–Marquardt is a popular numerical solution to find the smallest nonlinear function over parameter space. 

%The number of blocks available for performing the prediction divided in a ration of 70/15/15.

%\subsubsection{Input-Output mapping}
The following explains the input and output of the neural network model we use.

\begin{itemize}
    \item Input: Block values in the form of vector length, which indicate the number of previous values of the block time series.  The models without external input take a vector of the input  $y_i$ = \{$n_i$,  $s_i$\} while predicting the next blocks content either $n_i$ or $s_i$. 
    Similarly, the models with external input additionally take \{$x_i$\} as an input %where $x_i$=$\{f_i, Td_i, ms_i\}$ used as an external input 
    when the model is used to predict the subsequent blocks.

    \item Hidden layer: For  NAR and NARX, the number of hidden neurons is determined by performing a pre-analysis using the collected dataset. Based on this analysis, the models satisfy the Mean Square Error (MSE) value when the neurons are equal to ten. 
    
    \item The input delay $p$ and $q$ are approximated by using an autocorrelation $(p)$ and partial-autocorrelation $(q)$ plot. %Then it is used by both linear and nonlinear models. 
    
    \item Output: The predicted blocks $({y}^+_{i}|T_i=t)$ or $ {y}^+_{i}$)$ $ contain the predicted values of the blocks \{$n_{i}$, $s_{i}$\} of the weekend, working, and the combinations. 
\end{itemize}

% The Root Mean Square Error (RMSE) and Mean average error (MAE) metrics used to determine the models performance \cite{perfMetrics}. 

\subsection{Decision tree based classification }
%Miners play a crucial role in validating and generating blocks in the bitcoin system. 
Starting in 2010, there are more than 23 mining pools worldwide, as reported in Fig.~\ref{minersList}. It has been illustrated that mining pools compete to find the mathematical puzzle and the mining behavior is a game  \cite{bitcoinGame}\cite{bitcoinGameCorr}. %In the previous Section \ref{sec-dat}, it is illustrated that some of the major mining pools generate a block that may have some attributes that makes it more recognizable.  These block-related attributes are affected by the block generation strategy imposed by the mining pool, which means they may provide a mechanism to detect which of the mining pools are detectable. This paper observation shows that some feature sets can easily classify one of the leading mining pools.  

In this paper, we investigate if the mining pools are detectable using a machine learning, decision tree based  approach \cite{trees}\cite{Dtree}\cite{DctreeAnal}. 
 %Decision tree \cite{trees}\cite{Dtree}\cite{DctreeAnal} is a tree structure similar to the flow chart. 
It has a tree structure: Each branch represents the outcome of the test, and each leaf node represents a class label.  In some cases, it is essential to combine several decision trees to produce a better classification performance.  Such a combination produces an ensemble of different methods. In the present work, we considered two methods booted and RSUbooted \cite{matlab}. %, the other possible methods are not producing a better result. 

The accuracy, area under curve (AUC), sensitivity, and miss rate are used to test the  classification performance, in addition to false negative rate (FN), true positive rate (TP), and receiver operating characteristic (ROC) curve of TP versus TN, as commonly used for machine learning based classification \cite{perfMetrics}.  %Additionally, a false negative rate (FN) indicates that the current classifier assigns some percentage incorrectly to the positive class.  On the other hand, the true positive rate (TP) indicates that the current classifier assigns some percentages of the observations correctly to the positive class.  The receiver operating characteristic (ROC) curve shows a TP versus TN for the currently selected trained classifier. 
%Unlike the previous prediction part, which divided the dataset into a 70/15/15 ratio, in this part, we used $k$-fold validations, where $k=5$ gets better performance than the common $k=10$ values. 
\begin{figure*}[th!]
    \centering
    \subfigure[Empirical CDF of $s_i$]
    {
    \includegraphics[width=0.3\linewidth,height=0.27\linewidth]{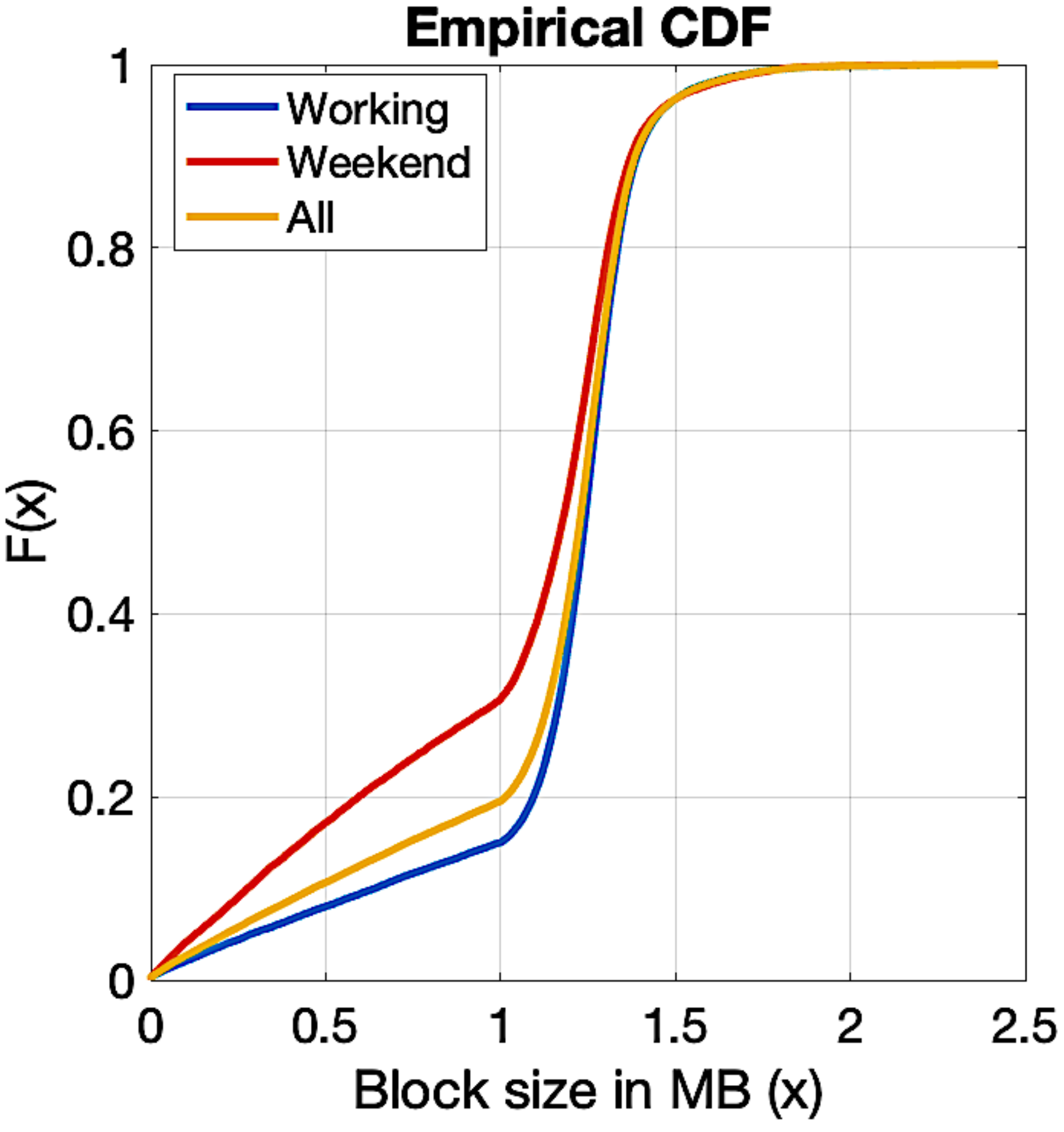}
   \label{bsize}
    }
    \subfigure[Empirical CDF of $n_i$]
    {
    \includegraphics[width=0.3\linewidth, height=0.27\linewidth]{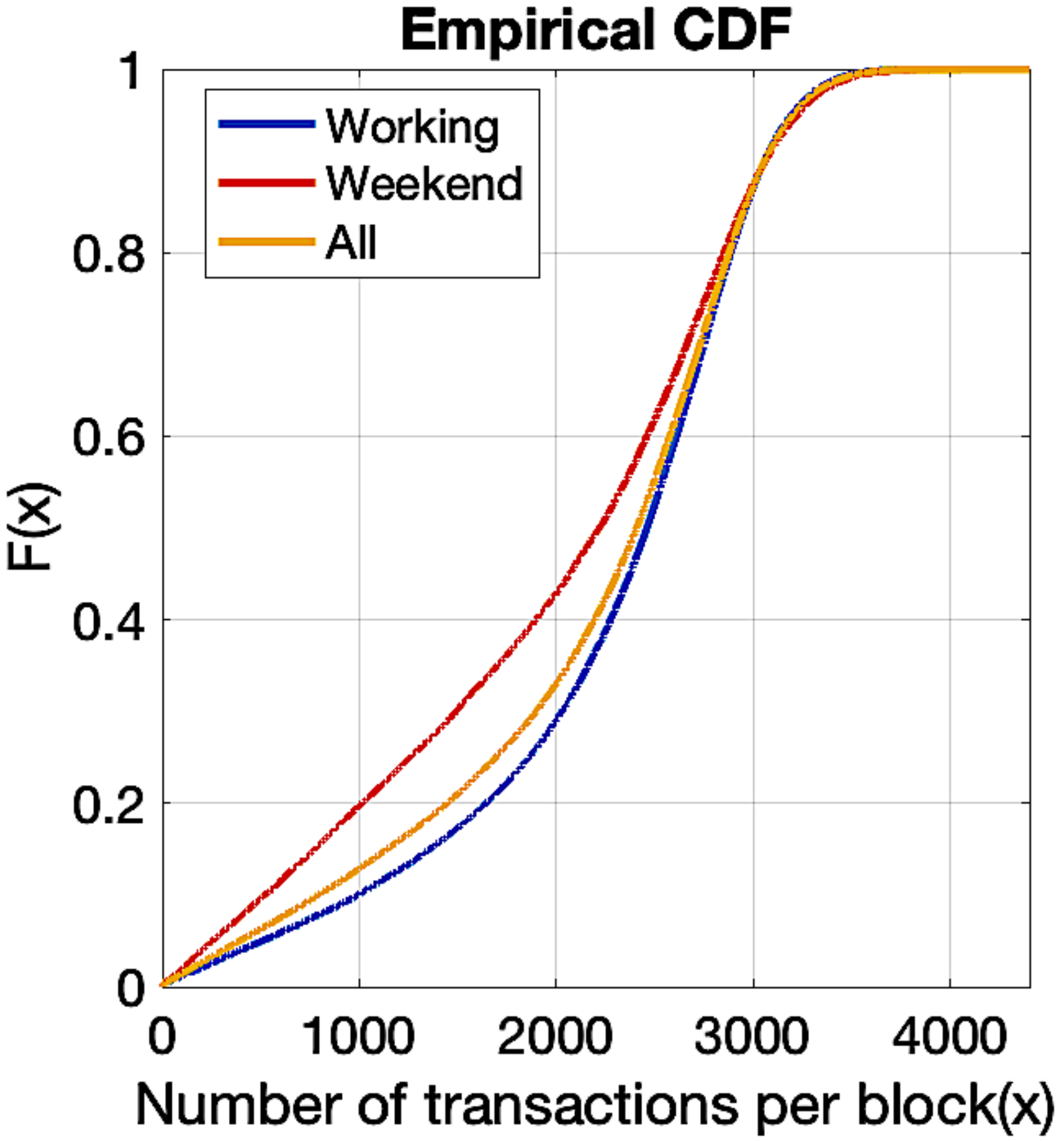}
   \label{nut}
    }
   % \label{kstest}
   \subfigure[Empirical CDF of $f_i$]
    {
    \includegraphics[width=0.3\linewidth, height=0.27\linewidth]{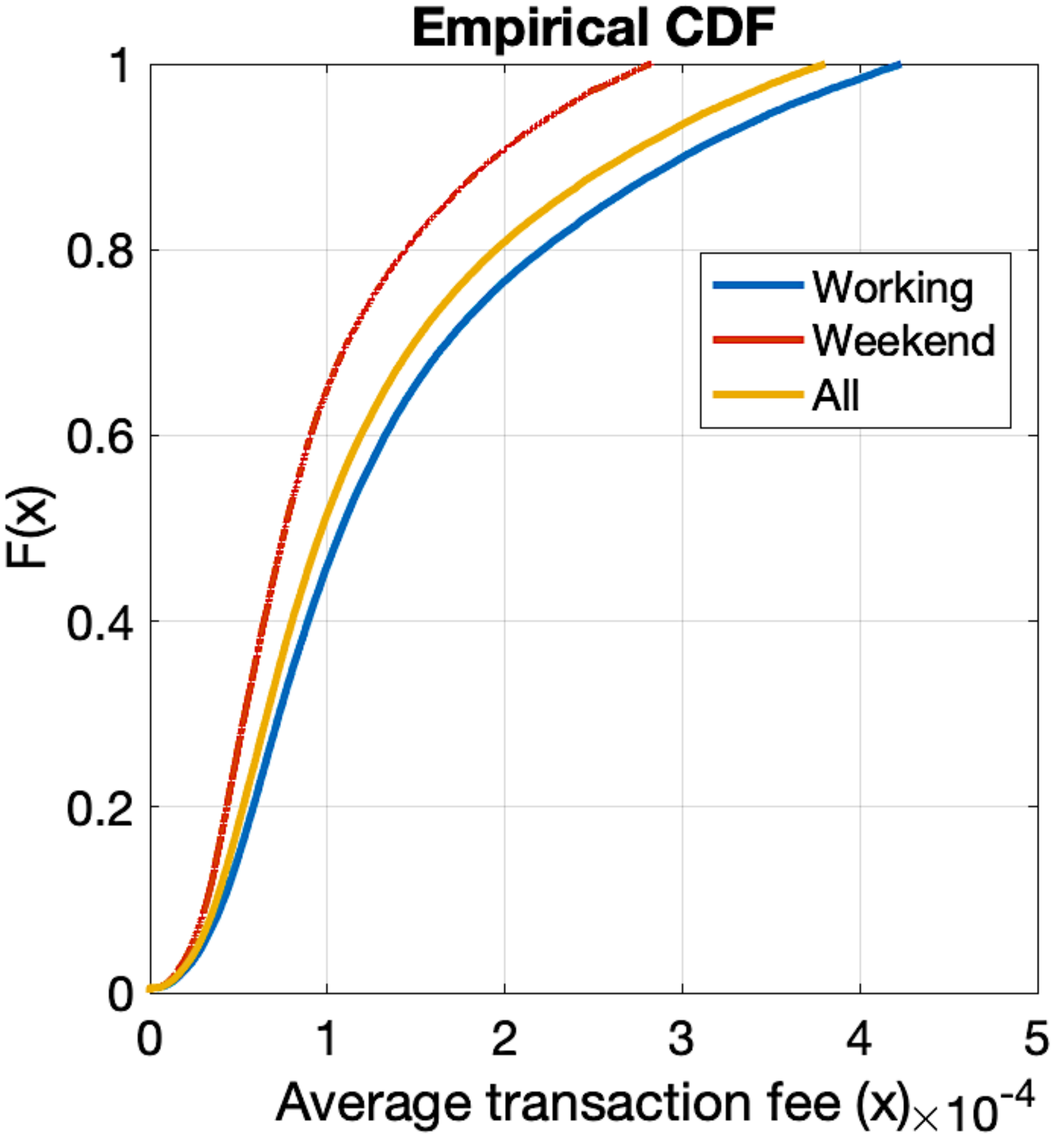}
   \label{averageFee}
    }
  \caption{CDFs of basic block attributes}%Empirical  
\end{figure*}

\section{Results: Exploratory Analysis}%{Bitcoin Transaction Handling: Statistical Analysis} 
\label{sec-sa}
%Statistical Properties of Bitcoin Transaction Handling Data

This section reports results and observations from an exploratory analysis of the collected data. 

\subsection{Basic block attributes}

The size $s_i$, the number of transactions $n_i$ and the fee $f_i$ are fundamental  attributes of a block. Since the transaction activities are time-varying process~\cite{TransBitcoin}, they may have distributions that vary from time to time, e.g.,  the weekend and working day demands have a different distribution.  

Fig.~\ref{bsize} reports in most cases (80\%), $s_i$ in working days has less than 1.4 MB, whereas it has a 1.2 MB size during weekends.  In both cases, the $s_i$ can grow more than 1.5 MB in 1\% of the cases.  Relatively, 30\% of the blocks have a size less than the default legacy size $s_i$ of 1 MB on weekend days; nevertheless, in working days, less than 20\% of the blocks have a size less than 1 MB. 

Similarly, Fig. \ref{nut} illustrates that the $n_i$ also varies as $s_i$: 50\% of the blocks have less than 2200 transactions per block in weekend days, while 2500 transactions per block in working days.  In working days, only 20\% of the blocks have $n_i$ less than 2100 transactions wherein the weekend, 40\% of the generated blocks have $n_i$ less than 2200. 
%\begin{figure}[ht!]
%\centering
 % \includegraphics[width=0.6\linewidth,height=0.44\linewidth]{AveragefeeCdf.pdf}
  %\caption{Empirical CDF of $f_i$}
  %\label{averageFee}
%\end{figure}

In addition, the miner's economic incentives affect which transactions to include in a block and this financial interest may also show some differences over time. %For instance, the $f_i$ per block varies on weekends and working days. 
Fig.~\ref{averageFee} reports, 50\% of the blocks, during weekdays, the $f_i$ is smaller than $1.3*10^{-4}$ while during working days, the $f_i$ is less than $1.43*10^{-4}$.  In both cases, 80\% of the $f_i$ is smaller than 0.00033 BTC, and with less than one percent, the $f_i$ can grow more significantly than 0.0004 BTC. % %%

\begin{table*}[ht]
\centering
 \caption{Major mining pools block related attributes properties}
%\small
\begin{tabular}{|l|l|l|l|l|}
\hline 
 Mining pool &  $\mu(s_i, n_i, f_i)$&  $\sigma(s_i, n_i, f_i)$& min($s_i, n_i, f_i$) &  max($s_i, n_i, f_i$)\\ \hline
 ?&(1.1252, $2.14*10^3$, $1.83*10^{-4}$)  &(0.3657, 844.2627, $2.18*10^{-4}$)  & ($2*10^{-4}$, 1, 0.00) & (2.4229,  4402, 0.0065)\\ \hline
 AntPool&(1.1141, $2.18*10^3$, $1.8*10^{-4}$)  &(0.3622, 844.2076, $1.9*10^{-4}$)  & ($3.34*10^{-4}$, 1, 0.00) & (2.2151, 4063, 0.0050) \\ \hline
 BTC.com& (1.0960, $2.15*10^3$, $1.86*10^{-4}$)  &(0.3782, 868.4394, $2.487*10^{-4}$)  & ($2.38*10^{-4}$, 1, 0.00) & (2.3056, 4243,  0.0121)\\ \hline
 F2Pool&(1.1099, $2.14*10^3$, $1.76*10^{-4}$)  &(0.3680, 845.6503, $2.16*10^{-4}$)  & ($2.66*10^{-4}$, 1, 0.00) & (2.3316, 4377, 0.0086)\\ \hline
 Poolin&(1.1091, $2.17*10^3$, $1.67*10^{-4}$)  &(0.3635, 842.1800, $1.87*10^{-4}$)  & ($2.17*10^{-4}$, 1, 0.00) & (2.3165, 3988,  0.0038)\\
 \hline
\end{tabular}
\label{majordistribution}
\end{table*}

\subsection{Miners}

\begin{figure}[th!]
\centering
  \includegraphics[width=0.9\linewidth,height=0.53\linewidth]{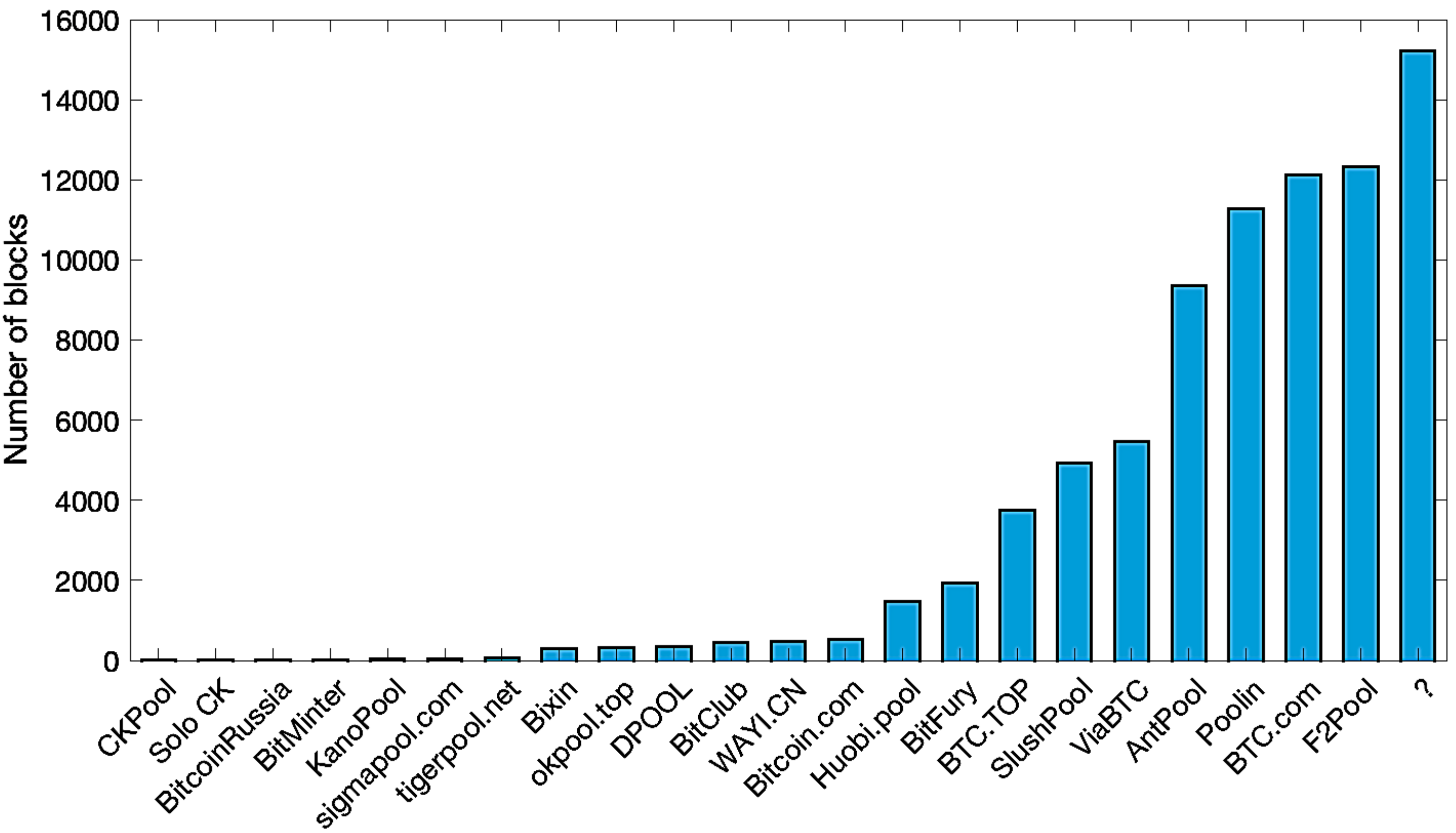}
  \vspace{-5pt}
  \caption{Miners}
  \label{minersList}
\end{figure}
\begin{figure}[th!]
\centering
  \includegraphics[width=0.9\linewidth,height=0.5\linewidth]{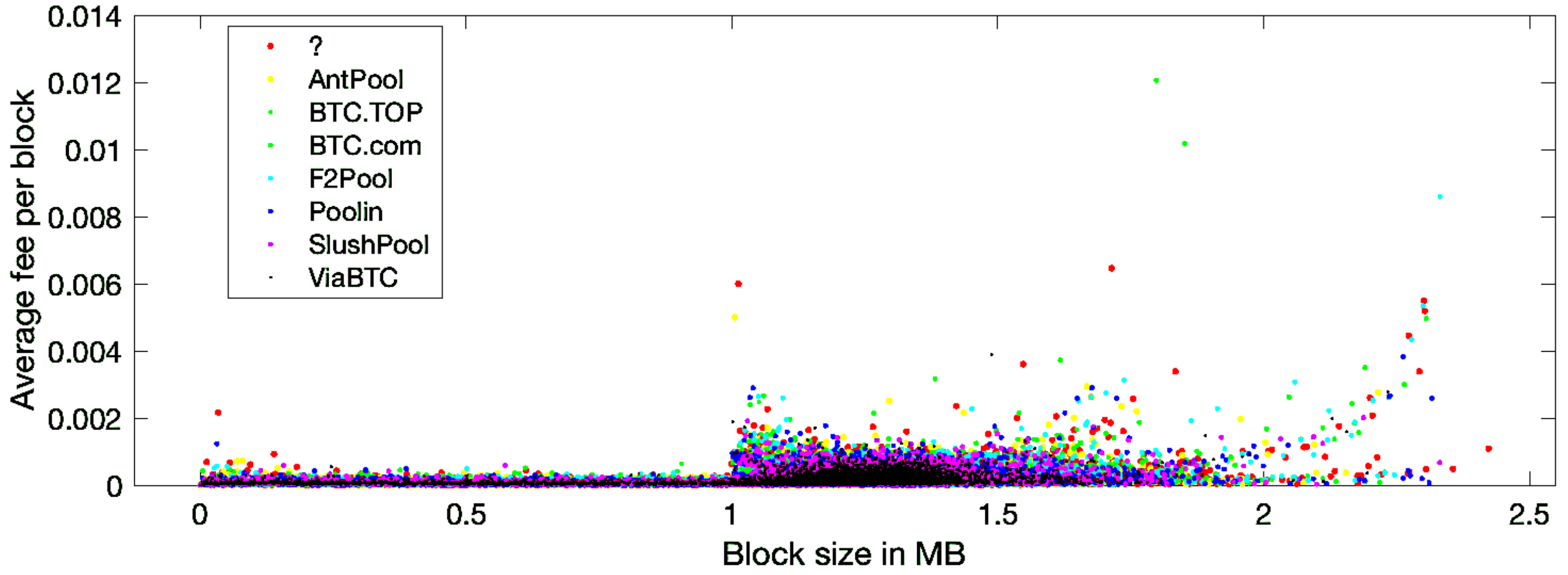}
  \vspace{-5pt}
  \caption{$f_i$ vs $s_i$}
  \label{fsize}
\end{figure}
\begin{figure}[th!]
\centering
  \includegraphics[width=0.9\linewidth,height=0.6\linewidth]{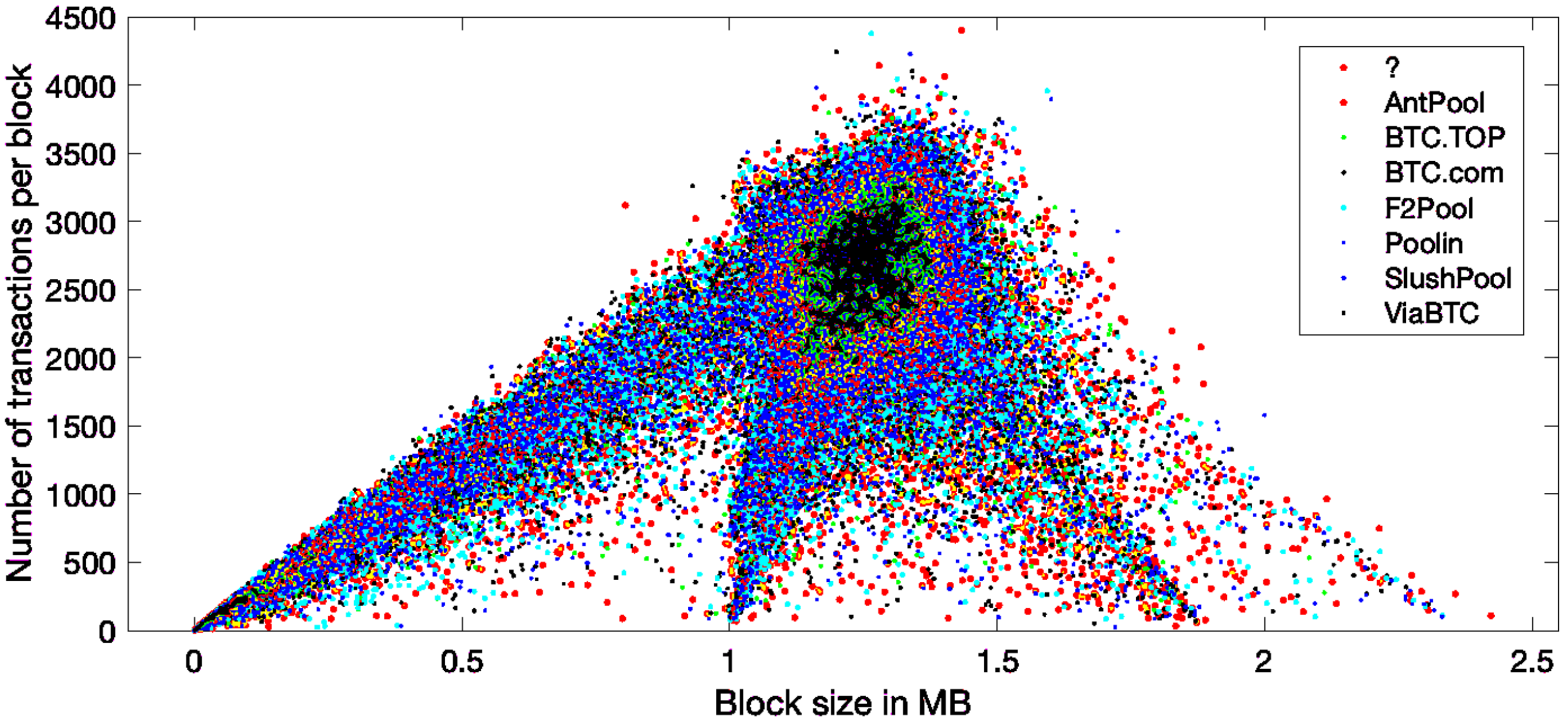}
  \vspace{-5pt}
  \caption{$n_i$ vs $s_i$}
  \label{nsize}
\end{figure}

Fig. \ref{minersList} reports miners' contribution in terms of the number of valid blocks in the main chain. As we can observe from the figure, unknown(?), F2Pool, BTC.com, Poolin, and AntPool contribute a higher number of blocks.  Combined, these five major mining pools generate around 50\% of the valid blocks. 

%It is natural for the miner to choose transactions with a higher fee to maximize the benefit. With miners' financial interest playing a significant role in choosing transactions from the backlog, miners or mining pools may develop a strategy that enhances the gain.  Perhaps, this miner or mining pool's behavior may also differ from working and weekend days.

Driven by the financial interest, a mining pool might use a strategy to increase the financial gain % and distribute the rewards to participating miners fairly 
\cite{majority}.  To explore, we analyze the blocks generated by the major mining pools.  Fig. \ref{fsize} and \ref{nsize} report that when the size $s_i$ is greater than 1.5 MB, it is visible from the figures that some of the major mining pools become more recognizable. However, when $s_i$ is less than 1 MB, it is challenging to see any difference between the pools.  Similarly, When the $s_i$ is between 1- 1.5 MB, we can see a high concentration of the mining pools. The figures also report that as the $s_i$ increases, the $n_i$ and $f_i$ also rise together.

 \begin{figure}[ht!]
\centering
  \includegraphics[width=0.8\linewidth,height=0.4\linewidth]{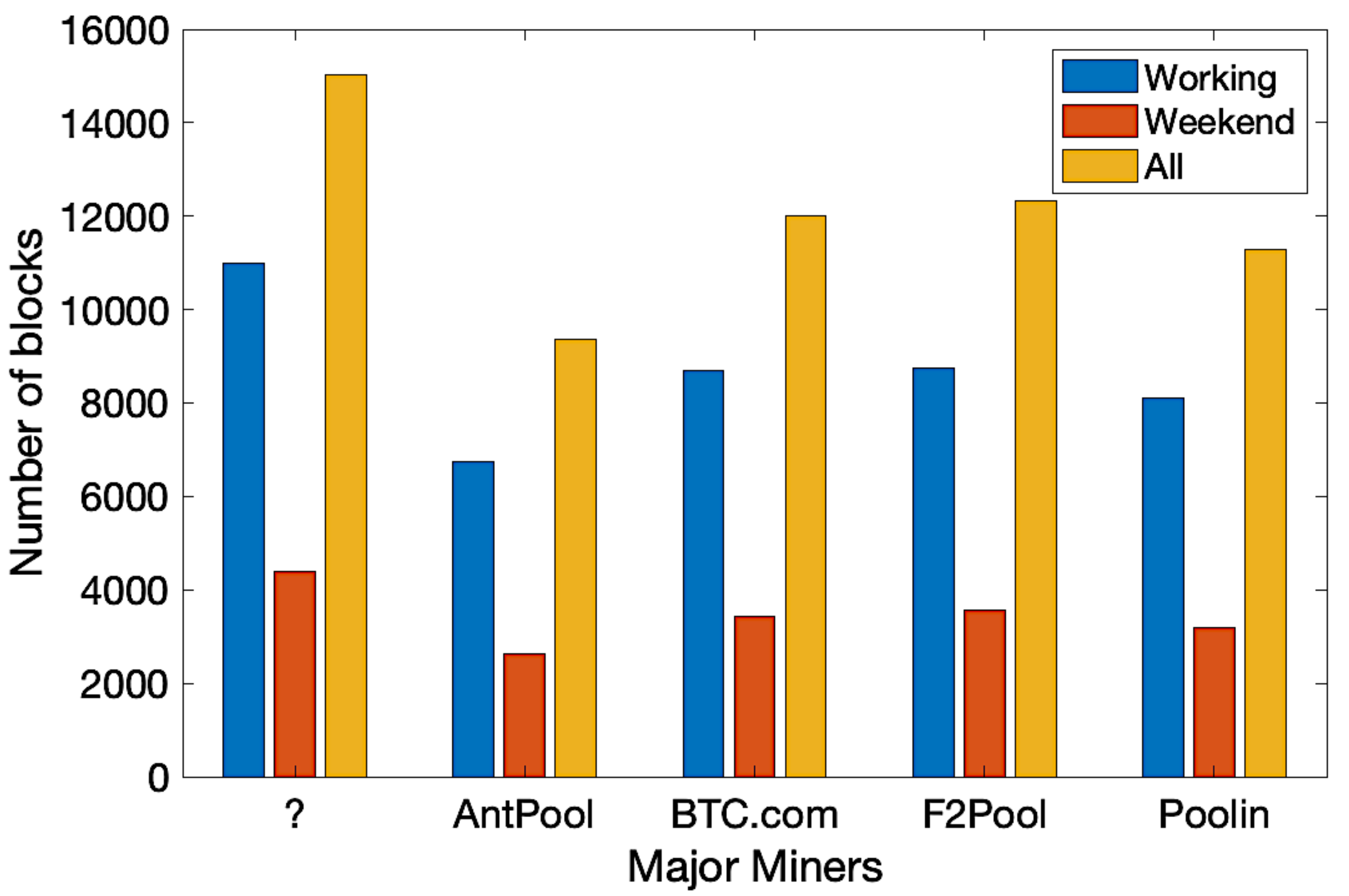}
  \caption{Miners contribution}
  \vspace{-5pt}
  \label{MinerContr}
\end{figure}

To further investigate the number of block contributions in the working and weekend days, we focus on the five major miners.  Fig. \ref{MinerContr} illustrates that these miners contribute similar number of blocks in the working days, except for the unknown(?) pool.  The same observation is also found for the weekend days. The unknown(?) pool generates a higher number of blocks in all cases.%, essentially, because it contributes more valid blocks than the rest. %in the weekend and working, and

To gain a deeper insight into the block contents than the number of blocks, Table \ref{majordistribution} is presented, where the mean $\mu$, standard deviation $\sigma$, minimum and maximum values of the basic block attributes ($s_i, n_i, f_i$) are shown. 
%illustrates the block-related features set of the mining pools properties.  The first column represents the five major mining pools based on the main chain's valid block contribution.  Similarly, the first row is the essential statistical measurement (mean ($\mu$), standard deviation ($\sigma$), min, max) of block-related attributes ($s_i, n_i, f_i$). 
Note that these major mining pools become operational starting 2016, except for Pooling in 2018 \cite{miningview}. Even though there is a gap in years between Poolin and the rest, Table \ref{majordistribution} shows that Poolin, F2Pool, BTC.com generate blocks with similar average size, %close to 1.09 MB,  
standard deviation and max values. 
However, the unknown (?) and AntPool generates block with size greater than the  three.  The unknown (?) has a block size  mean close to 1.214 MB, and the maximum block size is also found in this mining pool.  Additionally, the public mining pool, Poolin, comparing the maximum values of $f_i$ and $n_i$, has the smallest than the other four mining pools.

\subsection{Block generation}%{Inter-block generation time } %($Td_i$)

\subsubsection{Distribution of inter-block generation time} 
\begin{figure}[t!]
    \centering
    
    \subfigure[Inter-block generation]
    {
    \includegraphics[width=0.45\linewidth,height=0.23\linewidth]{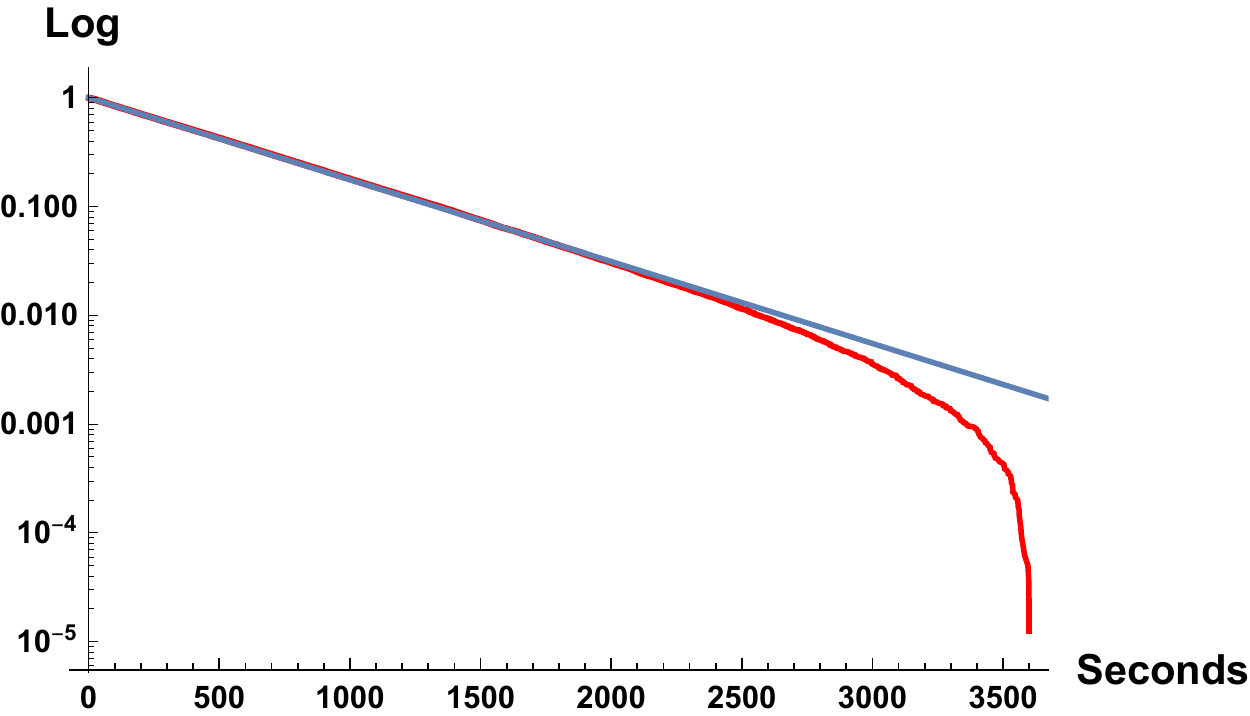}
   \label{inter}
    }
    \subfigure[Autocorrelation plot]
    {
    \includegraphics[width=0.45\linewidth, height=0.23\linewidth]{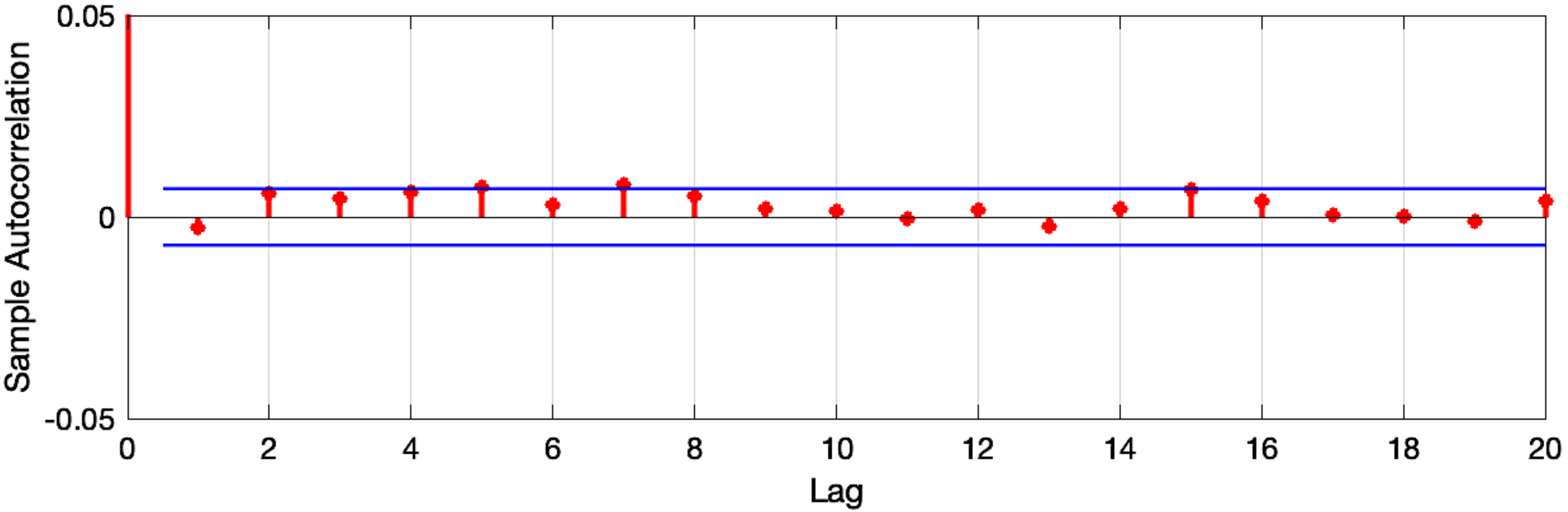}
   \label{auto}
    }
   % \label{kstest}
  \caption{Fitting of inter-block generation time to n.e.d}  
\end{figure}

Based on the Bitcoin design \cite{Nakamoto}, it has been expected that the inter-block generation time follows an exponential distribution, and the validity has also been checked~\cite{TransBitcoin}. Along the same line, Fig. \ref{inter} 
reports the fitting of the inter-block generation time to an exponential distribution. Additionally,  to check the independence of block generation time, its autocorrelation plot is illustrated in Fig. \ref{auto}.  As can be seen from Fig. \ref{inter} and Fig. \ref{auto}, the inter-block generation time fits well with an exponential distribution with increasing mismatch at the tail, partly due to the limited number of blocks in the dataset, and the autocorrelation is close to zero under all the lags in the figure, with the most significant difference only around 1\%, indicating that block generation is little correlated. 

\begin{figure}[th!]
    \centering
    \subfigure[100 minutes; $\lambda=9.44707$]
    {
    \includegraphics[width=0.43\linewidth,height=0.3\linewidth]{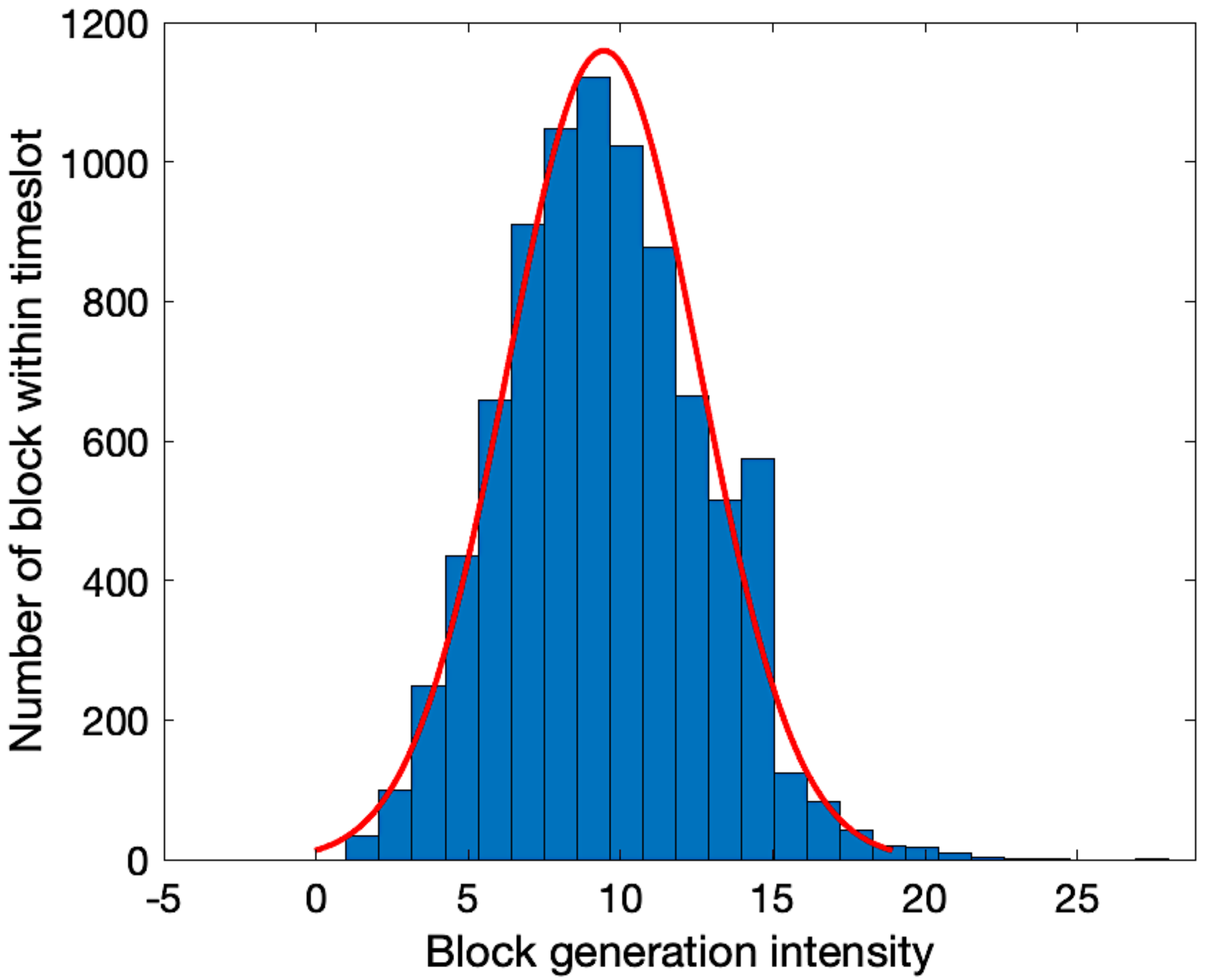}
   \label{inte100}
    }
    \subfigure[1000 minutes; $\lambda=103.184$] %Block generation intensity $\lambda=103.184$ (1000 minutes fixed time slot)
    {
    \includegraphics[width=0.47\linewidth, height=0.3\linewidth]{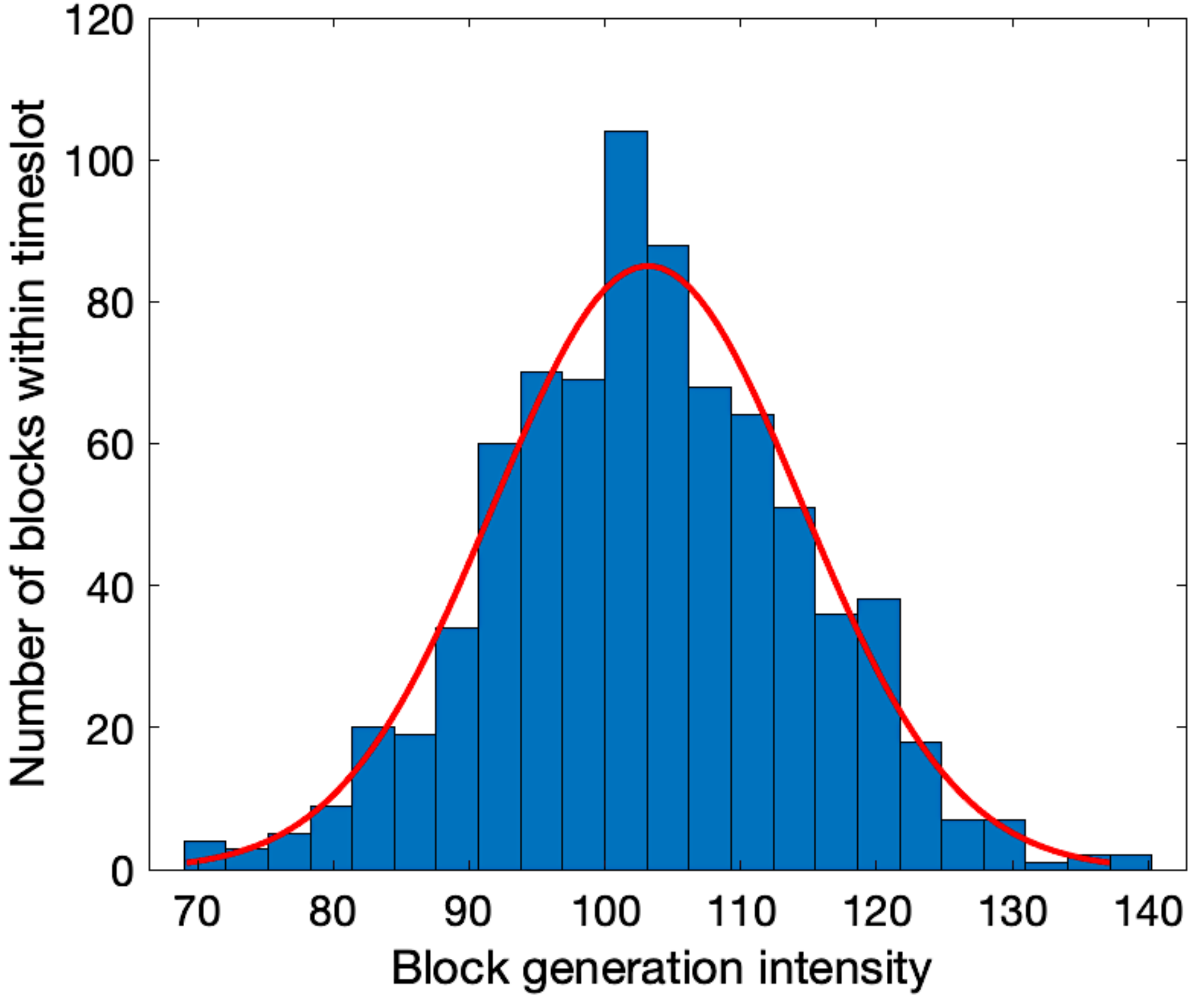}
   \label{inter1000}
    }
   % \label{kstest}
  \caption{Block generation histogram fitting to a Poisson distribution with intensity $\lambda$ under different time slot length}  
\end{figure}

\subsubsection{Fitting to a Poisson process}%{Block generation intensity}
Since the block generation process has exponentially distributed inter-generation times, we investigate if it can also be further treated as a Poisson process. For this, we make histograms of the number of blocks generated in different length of time and fit them with Poisson distributions. If the process is Poisson, these Poisson distributions must have the same intensity. For this investigation, Fig. \ref{inte100} and Fig. \ref{inter1000} are presented, where the best fitting intensity of Poisson distribution is shown under two time lengths, 100 and 1000 minutes. Clearly, the obtained two intensities differ noticeably, after taking into consideration that there is 10x scaling difference. This observation, which is surprising to us, implies that block generations can at most be approximated but cannot be treated as a Poisson process. 

%Fig. \ref{inte100} and Fig. \ref{inter1000} show that the block generation intensity within a fixed time slot of 100 and 1000 minutes do not conform very well to the Poisson distribution. In particular, the deviation between block generation intensity and the Poisson distribution seems to be most significant in the lower left-hand corner of the graph, which corresponds to the left tail of the Poisson distribution. The discrepancy is also noticeable in the upper right-hand corner of the graph, which corresponds to the right tail of the Poisson distribution.

%The graph shows that the smallest number of blocks within a fixed time slots are not small enough to be consistent with the Poisson distribution; similarly, the largest number of blocks within the time slots is not large enough to be consistent with the Poisson distribution.  The conclusion drawn from this is that the block generation intensity is not Poisson distributed, but this most probably comes from the timeslot size considered for the analysis. We further made an investigation through the chi-squared test. In both cases, the null hypothesis block generation tensity with 100 and 1000 minutes comes from Poisson distribution got rejected. 

\begin{figure}[t!]
\centering
  \includegraphics[width=0.9\linewidth,height=0.5\linewidth]{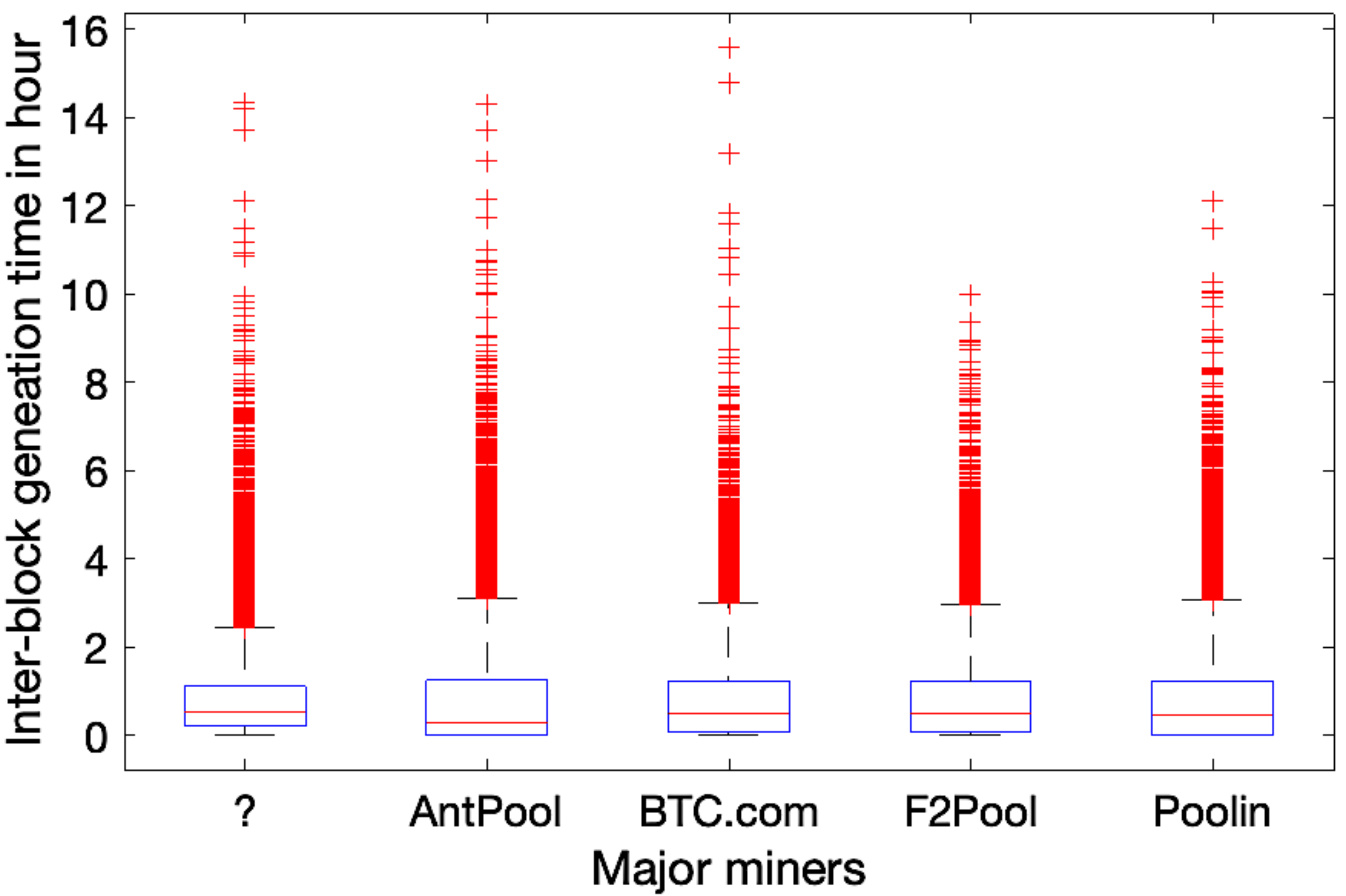}
  \caption{Major mining pools' inter-block generation time}
  \label{InterBlock}
\end{figure}

\subsubsection{Relation with miners}
To have a closer look on block generation, we made further investigation over the five major mining pools.  Fig. \ref{InterBlock} reports the inter-block generation time of the major mining pools. As the figure shows, the average inter-block generation time is almost the same among the major mining pools. However, for the median, there is visible difference: While for Unknown(?) and F2Pool, the median time is close to 52 minutes, for BTC.com and Poolin, it is near 45 minutes and for AntPool, it is close to half-hour.  The minimum inter-block generation time is the same for all major mining pools, close to zero. However, for the maximum inter-block generation time, while AntPool and Unknown(?) need 14 hours and 30 minutes, BTC.com demands 16 hours.  In addition, the public mining pool, Poolin, requires 12 hours, and  unlike or shorter than the others, F2Pool needs only 10 hours. As a highlight from Fig. \ref{InterBlock}, F2Pool stands clearly out of the others with shortest tail.

%(YM: I don't understand this paragraph.) As a highlight from Fig. \ref{InterBlock}, AntPool distributions seem to have positively skewed, which means the distribution occurs more on the one side of the scale with a long tail on the right side. Unknown(?), BTC.com, Poolin, and F2Pool distributions are normally distributed, reflecting that data near the mean are more frequent in occurrence than data far from the mean.  The inter-block generation by the Unknown (?) shows shorter intervals than the rest. This is perhaps because it contributes more valid blocks than the rest four mining pools. 

\subsubsection{Relation with basic block attributes}
%To understand how block generation may be affected by the basic attributes, 

\begin{figure}[th!]
    \centering
    
    \subfigure[Inter-block generations vs $s_i$]
    {
    \includegraphics[width=0.45\linewidth,height=0.45\linewidth]{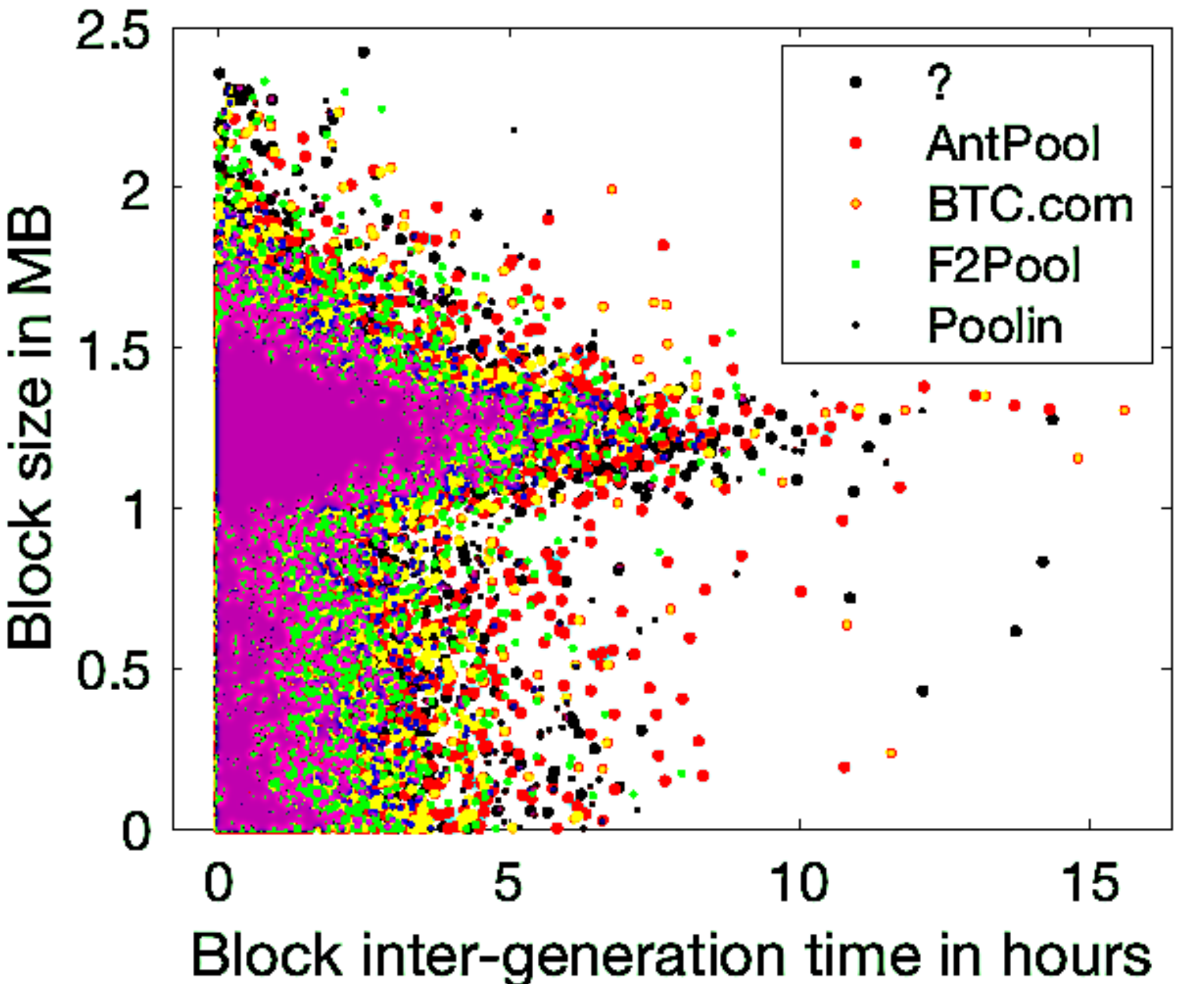}
   \label{Interbsize}
    }
    \subfigure[Inter-block generations vs $n_i$]
    {
    \includegraphics[width=0.45\linewidth, height=0.45\linewidth]{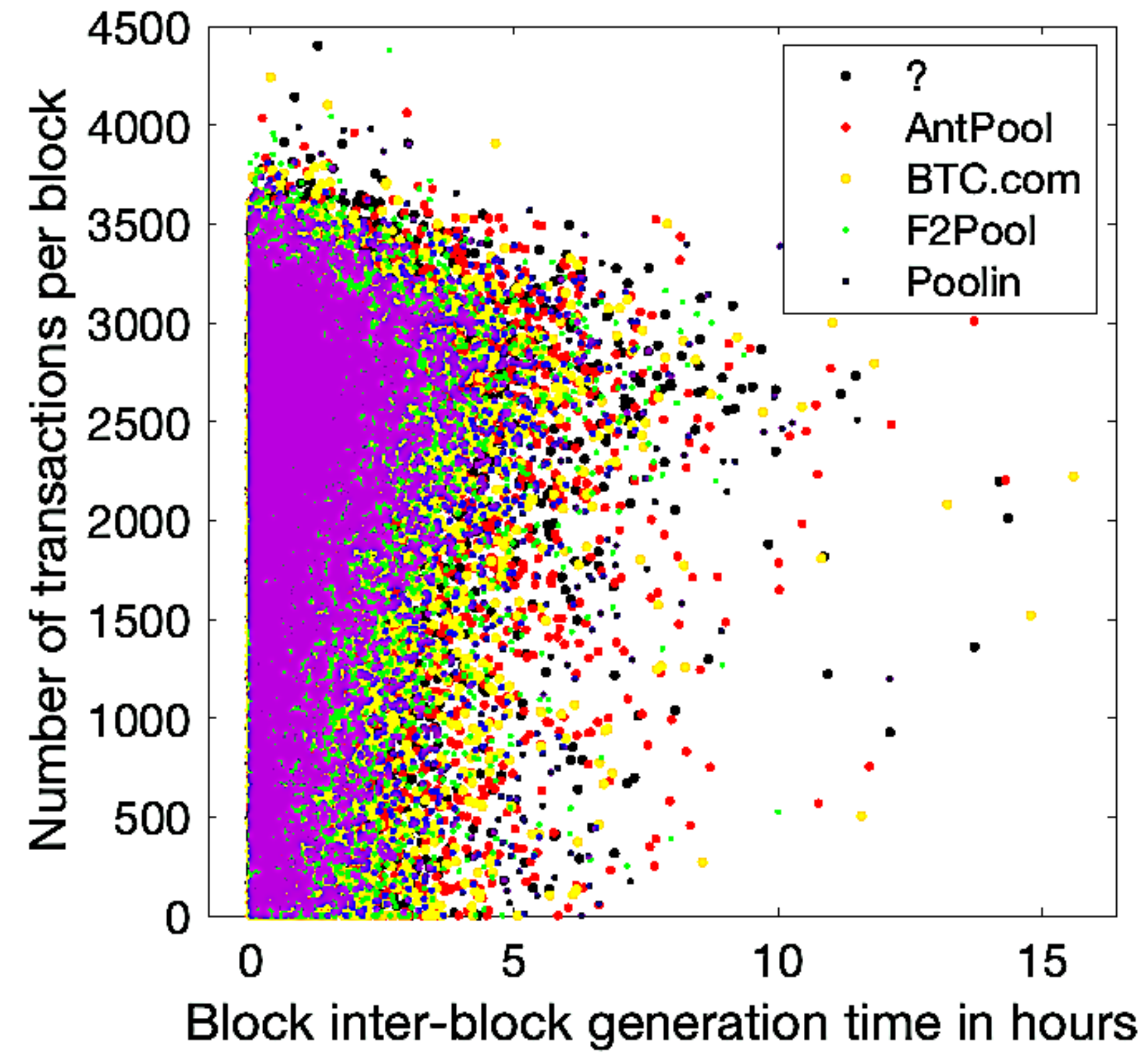}
   \label{Internut}
    }
    \subfigure[Inter-block generations vs $f_i$]
    {
    \includegraphics[width=0.45\linewidth, height=0.45\linewidth]{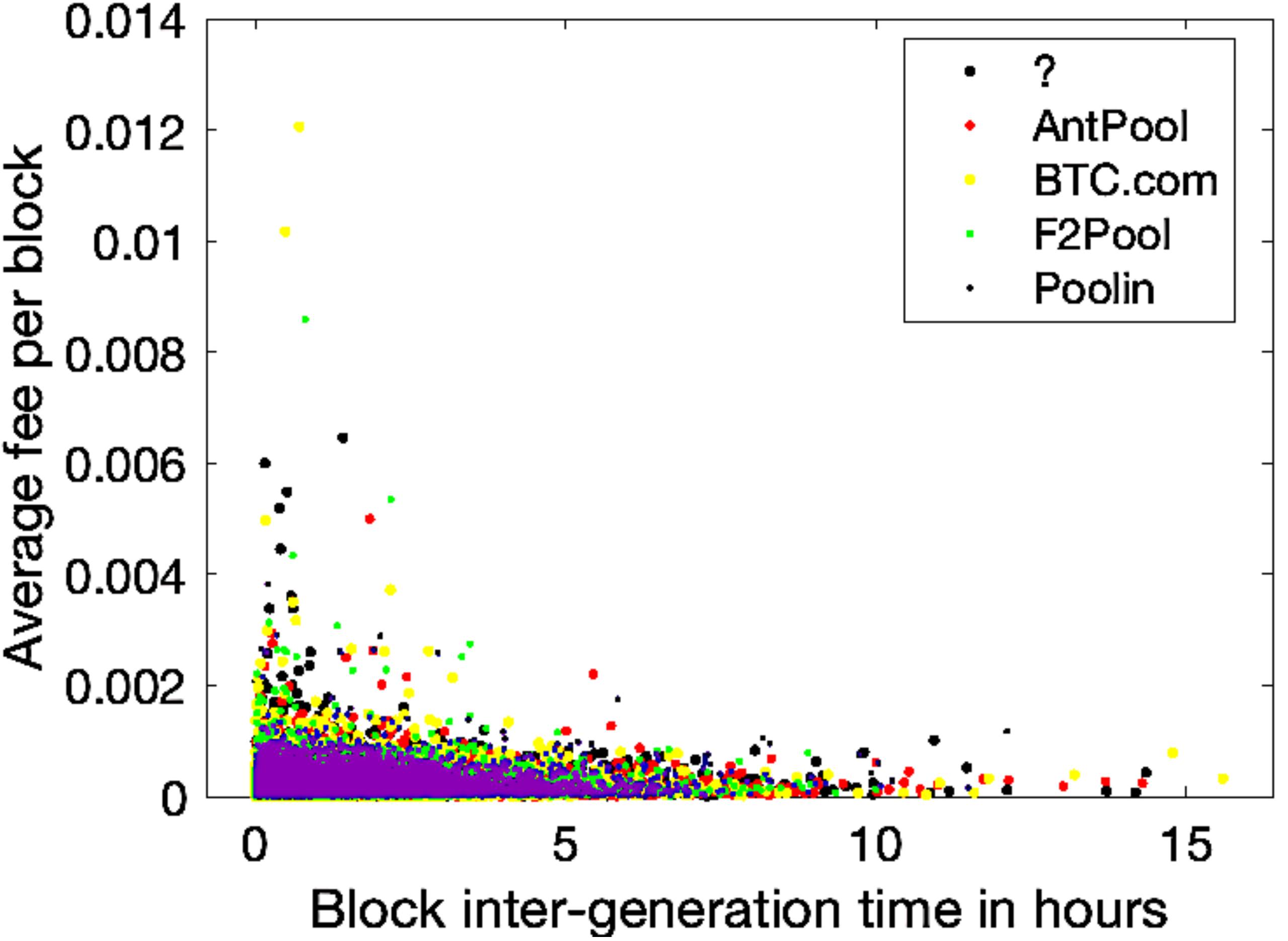}
   \label{Interfee}
    }
   % \label{kstest}
  \caption{Inter-block generations v.s. block size, transaction number and fee}  
\end{figure}

%Fig. \ref{MinerContr} shows the four major mining pools having a close number of blocks over weekends and working days.  In addition to this, Fig. \ref{InterBlock} reports the mining pools' inter-block generation time of the top pools. 
We further explored the relationship  between block generation and the three basic block attributes, shown by Fig. \ref{Interbsize}, \ref{Internut}, and \ref{Interfee}. Specifically,  Fig. \ref{Interbsize} illustrates that when the block size $s_i$ is greater than 1.5 MB, the inter-block generation time seen by the blocks is less than two hours.  However, when the block size is concentrated between 1-1.5 MB, Unknown(?), AntPool, and BTC.com block can have the inter-block generation time greater than 13 hours.  On the other hand, the blocks from Poolin and F2Pool seem to be generated with shorter interval than the rest three, which is also indicated by Fig. \ref{InterBlock}.

% \begin{figure}[ht!]
%\centering
  %\includegraphics[width=0.45\linewidth,height=0.45\linewidth]{InterBlocksize.pdf}
  %\caption{Miners inter-block generations vs $s_i$}
 % \label{Interbsize}
%\end{figure}

% \begin{figure}[ht!]
%\centering
 % \includegraphics[width=0.45\linewidth,height=0.45\linewidth]{InterNut.pdf}
 % \caption{Miners inter-block generations vs $n_i$}
  %\label{Internut}
%\end{figure}
% \begin{figure}[ht!]
%\centering
 % \includegraphics[width=0.9\linewidth,height=0.35\linewidth]{InterFee.pdf}
  %\caption{Miners inter-block generations vs $f_i$}
  %\label{Interfee}
%\end{figure}

In addition, Fig. \ref{Internut} demonstrates that the number of transactions $n_i$ in a block of Poolin  is on average smaller than the other mining pools.  Most of the $n_i$ from F2Pool seem to have a shorter inter-block generation time. However, it is hard to say for the Unknown(?) and AntPool, because the plot shows most of the block with $n_i$ seems to have a larger inter-block generations time. These effects may arise from the state of the mempool, when the mempool contains more transactions then the miners can pick as much number of transaction to include in block. 

Furthermore, it is natural the miners prioritize the finical incentives, which encourages the miners to pick up transactions with a higher fee. Fig. \ref{Interfee} illustrates this fact. Specifically,  when the fee $f_i$ is higher, the inter-block generation time of the block is lower, maybe even shorter than an hour. The figure also shows that the blocks with the smaller average fee from Unknown(?), AntPool, and BTC.com may experience inter-block generation time greater than 14 hours.  On the other hand, the blocks from Poolin seem to have a less average fee and seeingly smaller inter-block generation time. %low block, 

\subsection{Transaction arrival and confirmation time}%{Transaction inter-arrival and confirmation time}
Users generate transactions for validation. New arrivals stay at the backlog (memory pool) until the nonce finding is successful and they are picked up by the miner. %These transactions propagate into the Bitcoin network.

\begin{figure}[tb!]
    \centering
    
    \subfigure[Transaction inter-arrival  time ]%fitting n.e.d
    {
    \includegraphics[width=0.45\linewidth,height=0.23\linewidth]{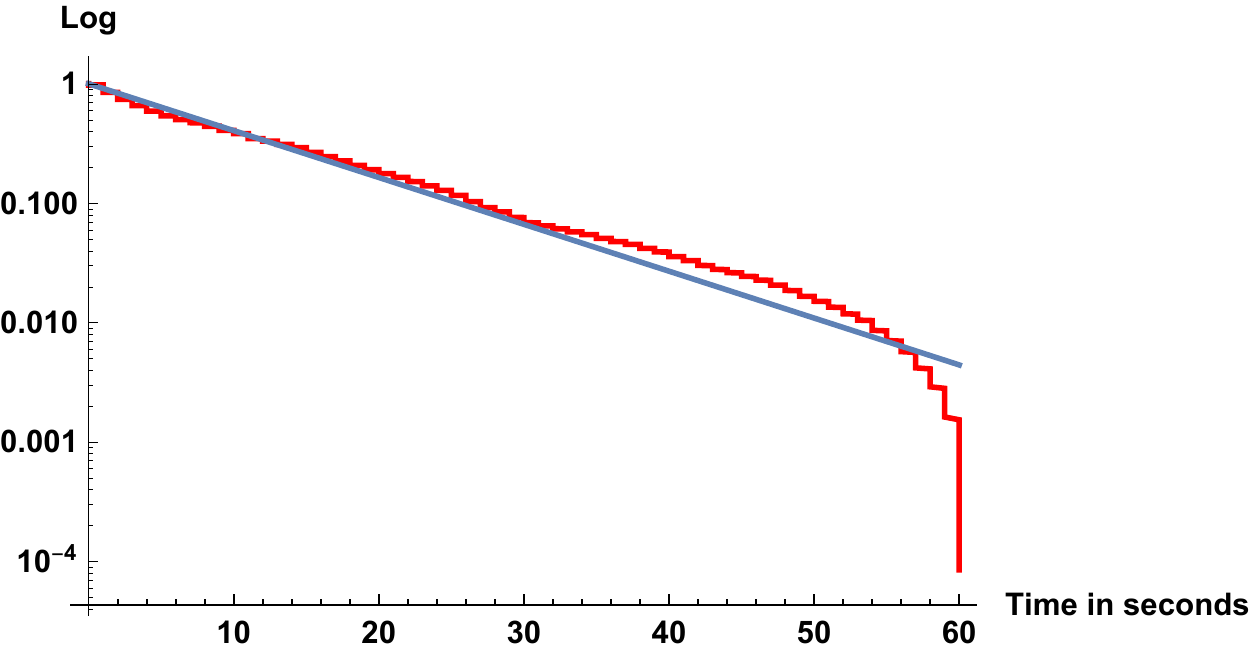}
   \label{transconinterarrival}
    }
    \subfigure[Autocorrelation plot]
    {
    \includegraphics[width=0.45\linewidth, height=0.23\linewidth]{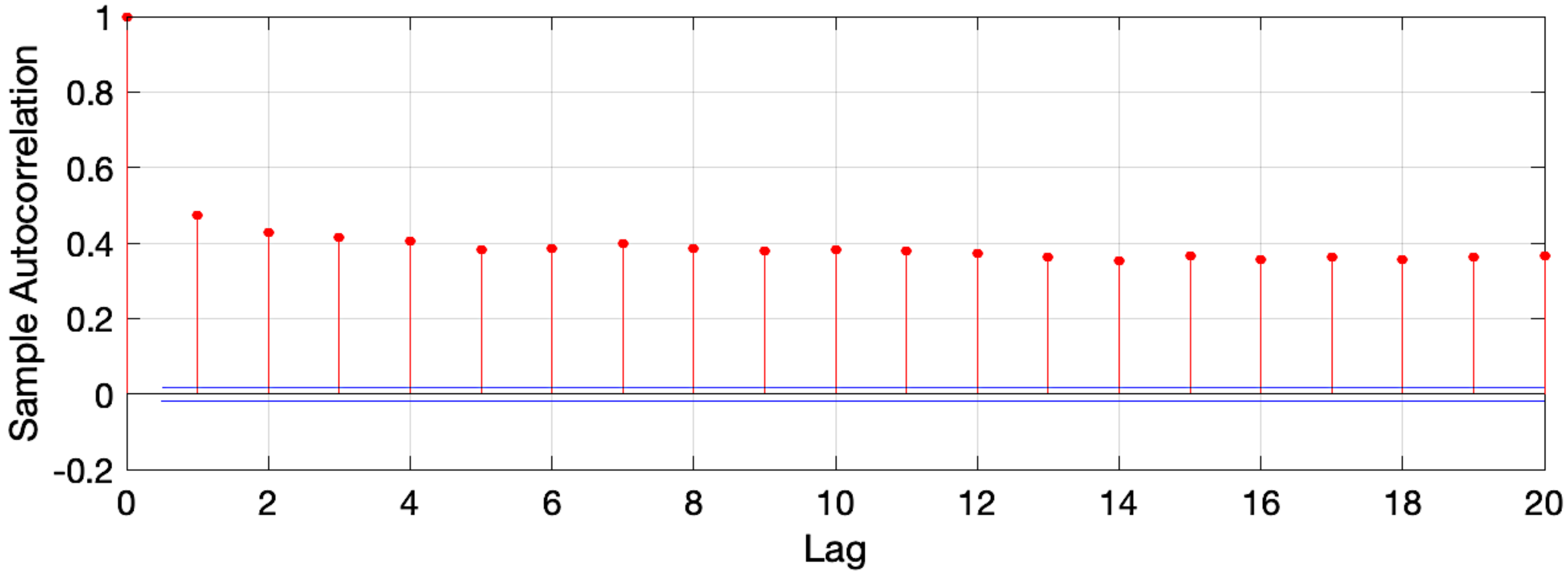}
   \label{autoarrival}
    }
   % \label{kstest}
  \caption{Transaction inter-arrival time fitting n.e.d}  
\end{figure}
\begin{figure}[th!]
    \centering
    
    \subfigure[Transaction confirmation time ]%fitting n.e.d
    {
    \includegraphics[width=0.45\linewidth,height=0.23\linewidth]{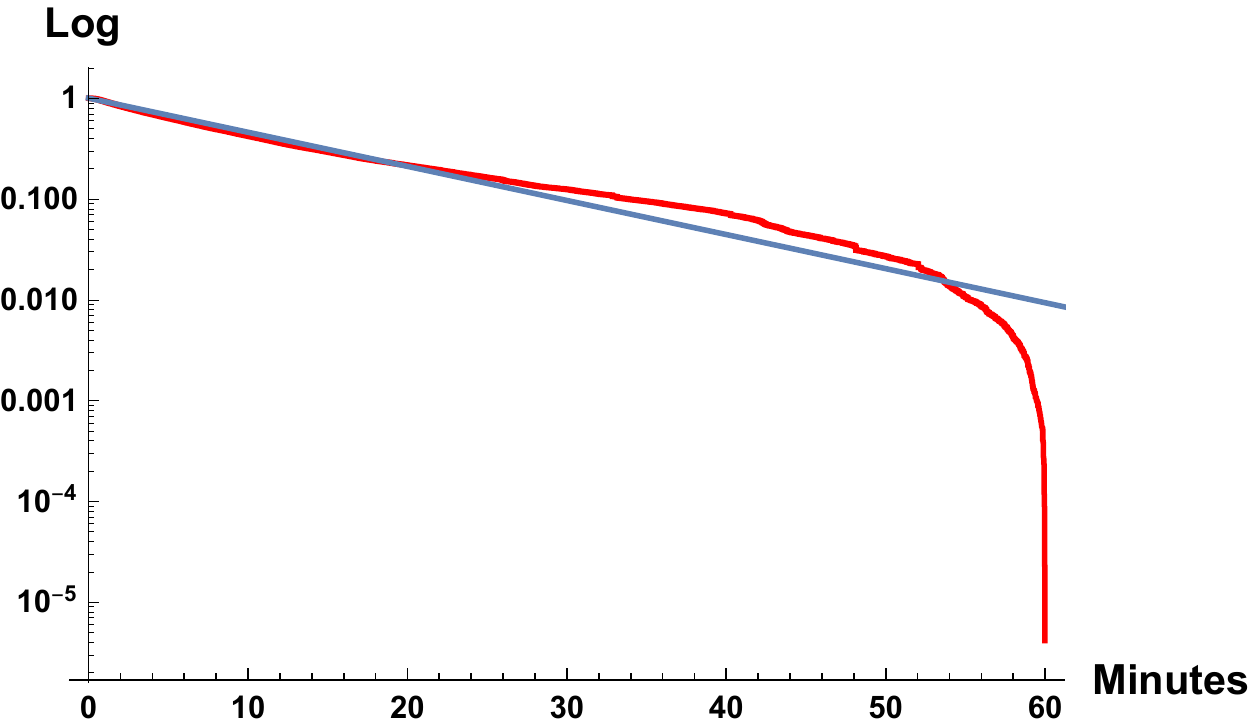}
   \label{transconfirmfit}
    }
    \subfigure[Autocorrelation plot]
    {
    \includegraphics[width=0.45\linewidth, height=0.23\linewidth]{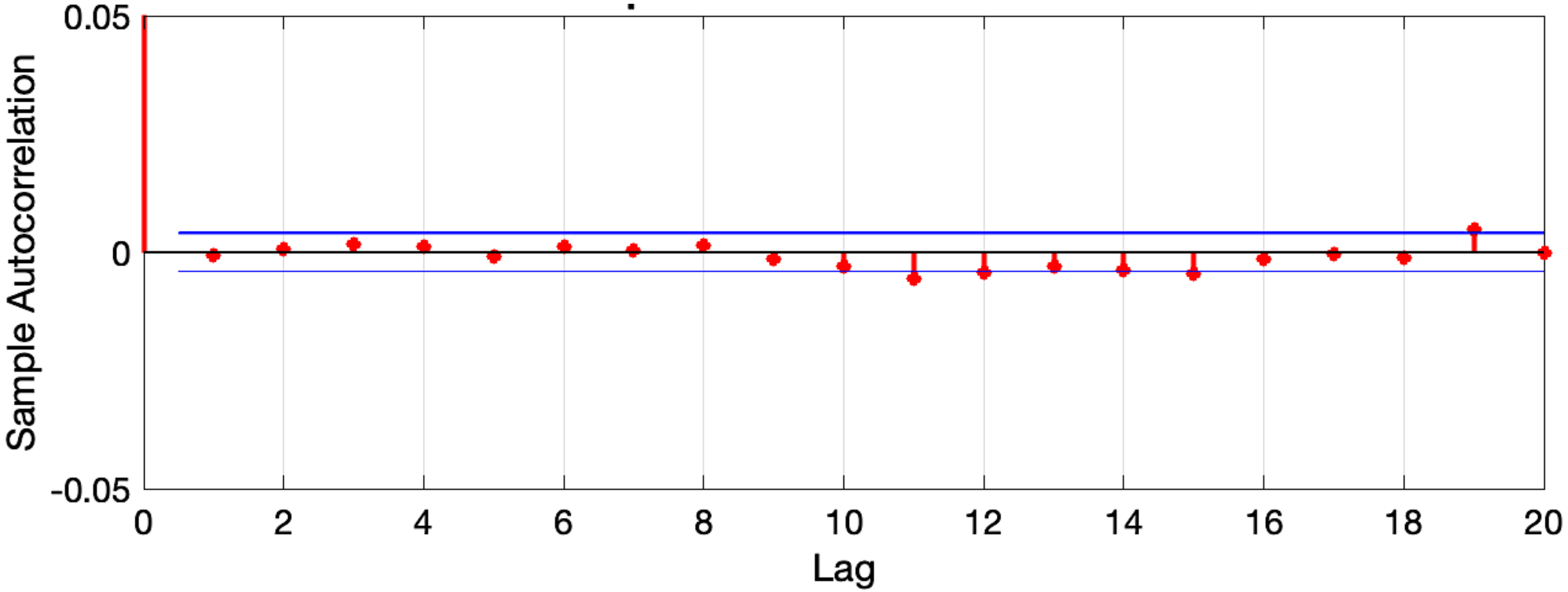}
   \label{autocorr}
    }
   % \label{kstest}
  \caption{Transaction confirmation time fitting n.e.d}  
\end{figure}

\subsubsection{Transaction inter-arrival  time}
%It is assumed that the transaction arrival process is a Poisson process in Bitcoin.  
Fig. \ref{transconinterarrival} shows that the fitting of transaction inter-arrival times to a negative exponential distribution is only reasonable well with visible deviation.  
Additionally, Fig \ref{autoarrival} reports the inter-arrival between the transactions is correlated. These reflect that there exists some level of  dependence between transaction arrivals. 

%\subsection{Transaction confirmation time}
\subsubsection{Transaction confirmation time}
Fig. \ref{transconfirmfit} reports the transaction confirmation time fitting to a negative exponential distribution, with a sharp drop at the tail. %^, likely attributed to the miner financial incentives. 
Additionally, Fig.  \ref{autocorr} illustrates that the transaction confirmation time is uncorrelated, reflecting that the transaction confirmation time is independent. 

\begin{figure}[t!]
   % \centering
    \subfigure[Transaction fee effect]
    {
    \includegraphics[width=0.45\linewidth, height=0.3\linewidth]{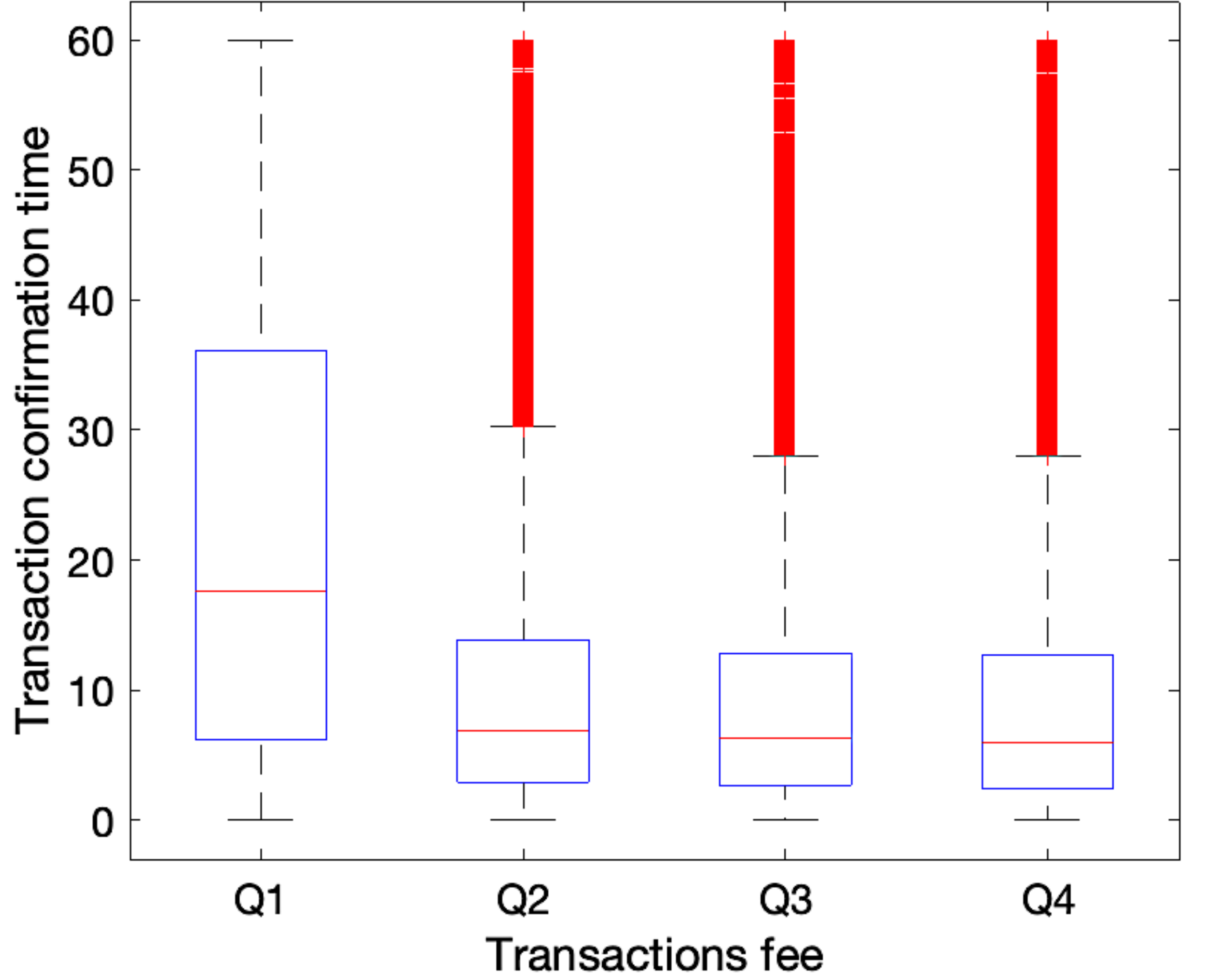}
   \label{waiting}
    }
    \subfigure[log plot]
    {
    \includegraphics[width=0.45\linewidth, height=0.3\linewidth]{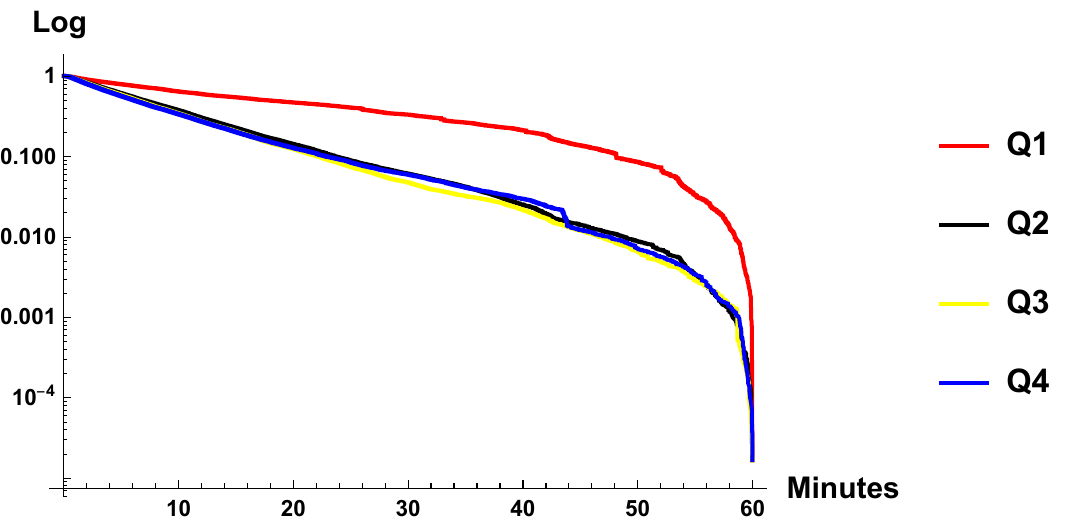}
   \label{feeeffec}
    }
   % \label{kstest}
  \caption{Transaction fee effect on transaction confirmation time}  
\end{figure}

Since a miner tends to choose transactions with a higher fee, to demonstrate this effect on the confirmation time, Fig .\ref{waiting} is presented.  Specifically, it demonstrates the relationship of confirmation time and fee for Q1 (25\%), Q2 (50\%), Q3 (75\%), and greater than Q3, i.e., (Q4) for $f_b$. Their intervals are respectively (0,Q1), (Q1,Q2), (Q2,Q3), and (Q4,$\infty$). As Fig \ref{waiting} shows, low fee transactions exhibit a higher confirmation time. On average, the low fee transactions (Q1) wait 22 minutes for validation. However, for higher fee (Q4) transactions, the average confirmation time is less than half of that of the low fee transactions. % confirmation time, 9 minutes and thirty seconds.  
For Q2 and Q3, the transactions exhibit close to a ten-minute average confirmation time. Still, transactions from Q2, on average, wait one more minute extra than Q3.  Overall, transactions wait on average 13 minutes, and we also observed a few transactions waiting for more than 24 hours at the backlog.  At the same, these few transactions also tend to have a fee associated relatively very small. 
 % \begin{figure}[ht!]
%\centering
 % \includegraphics[width=0.8\linewidth,height=0.6\linewidth]{TransactionWaiting.pdf}
  %\caption{Transaction confirmation time}
  %\label{transconfirm}
%\end{figure}

 %\begin{figure}[ht!]
%\centering
 % \includegraphics[width=0.9\linewidth,height=0.45\linewidth]{transactionWaiting.pdf}
  %\caption{Transaction confirmation time fitting n.e.d}
  %\label{transconfirmfit}
%\end{figure}
 %\begin{figure}[ht!]
%\centering
 % \includegraphics[width=0.9\linewidth,height=0.35\linewidth]{autoWait.pdf}
  %\caption{Autocorrelation }
 % \label{autocorr}
%\end{figure}

%\begin{figure}[ht!]
%\centering
 % \includegraphics[width=0.8\linewidth,height=0.43\linewidth]{transactionWaiting.pdf}
 % \caption{Transaction confirmation time fitting n.e.d}
 % \label{transconfirmfit}
%\end{figure}

%\begin{figure}[ht!]
%\centering
 % \includegraphics[width=0.8\linewidth,height=0.5\linewidth]{waiting.pdf}
  %\caption{Transaction fee effect on confirmation time}
 % \label{waiting}
%\end{figure}

%\begin{figure}[ht!]
%\centering
 % \includegraphics[width=0.8\linewidth,height=0.4\linewidth]{FeeEffect.pdf}
  %\caption{Transaction fee effect on confirmation time}
  %\label{feeeffec}
%\end{figure}

\section{Results: Predictability Analysis}\label{sec-pa}
%{Bitcoin Transaction Handling: Is It Predictable?}

Having explored the various characteristics of transaction handling in the previous section, this section is devoted to investigating if and what such characteristics can be predicted. For this predictability analysis, the prediction approaches introduced in Section \ref{sec-met} are used. The results are reported and discussed in the rest of this section, where the dataset is divided into three parts, i.e, training, test and validation, and the details of this division is reported in Table \ref{database}.

%\subsection{RESULTS AND DISCUSSION}\label{sec-res}
%Transactions vary over time \cite{TransBitcoin}.  The $s_i$ and $n_i$ also vary over time, which makes throughput vary from the weekend to the working days. Additionally, the mempool state is dependent on transaction arrival and block size. Based on these facts, we divided the dataset into working and weekends accordingly, as reported in Table \ref{database}.
\begin{table}[t!]
\centering
%\tiny
 \caption{Division of the dataset} %Database size and blocks division
\begin{tabular}{|p{16mm}|p{10mm}|p{10mm}|p{10mm}|p{16mm}|}
\hline
Dataset &  Training  & Test & Validation & \#No of blocks \\
\hline
Working\_day &  40095  & 8591 & 8591 & 57277\\
\hline
Weekend\_day &  16190  & 3469 & 3469 & 23128\\
 \hline
 All\_db & 56286 & 12061  &  12061 & 80408\\
 \hline
\end{tabular}
 \label{database}
\end{table}

%In this section,  the result from two cases is demonstrated: Firstly, linear and nonlinear models' ability to predict the Bitcoin block values. Then, secondly, decision-tree models performance to detect major mining pools block generation behavior.

\subsection{Basic block attributes}%{Throughput prediction}
\begin{figure}[ht]
    \centering
    \subfigure[measured vs predicted $s_i$ ]{
    \includegraphics[width=0.9\linewidth,height=0.5\linewidth]{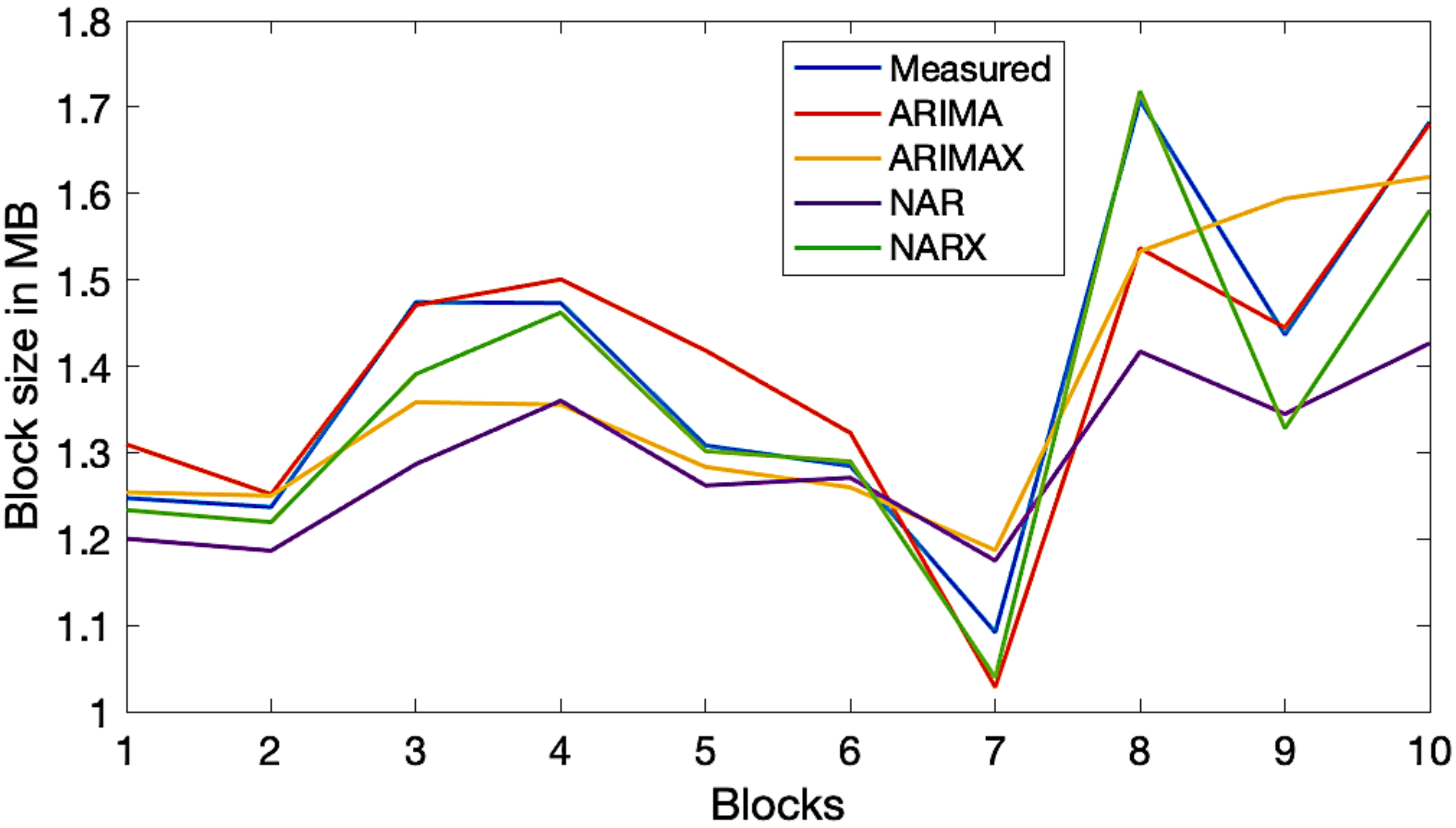}
   \label{weekendbsize}
    }
   
    \subfigure[measured  vs predicted $n_i$ ]{
    \includegraphics[width=0.9\linewidth,height=0.5\linewidth]{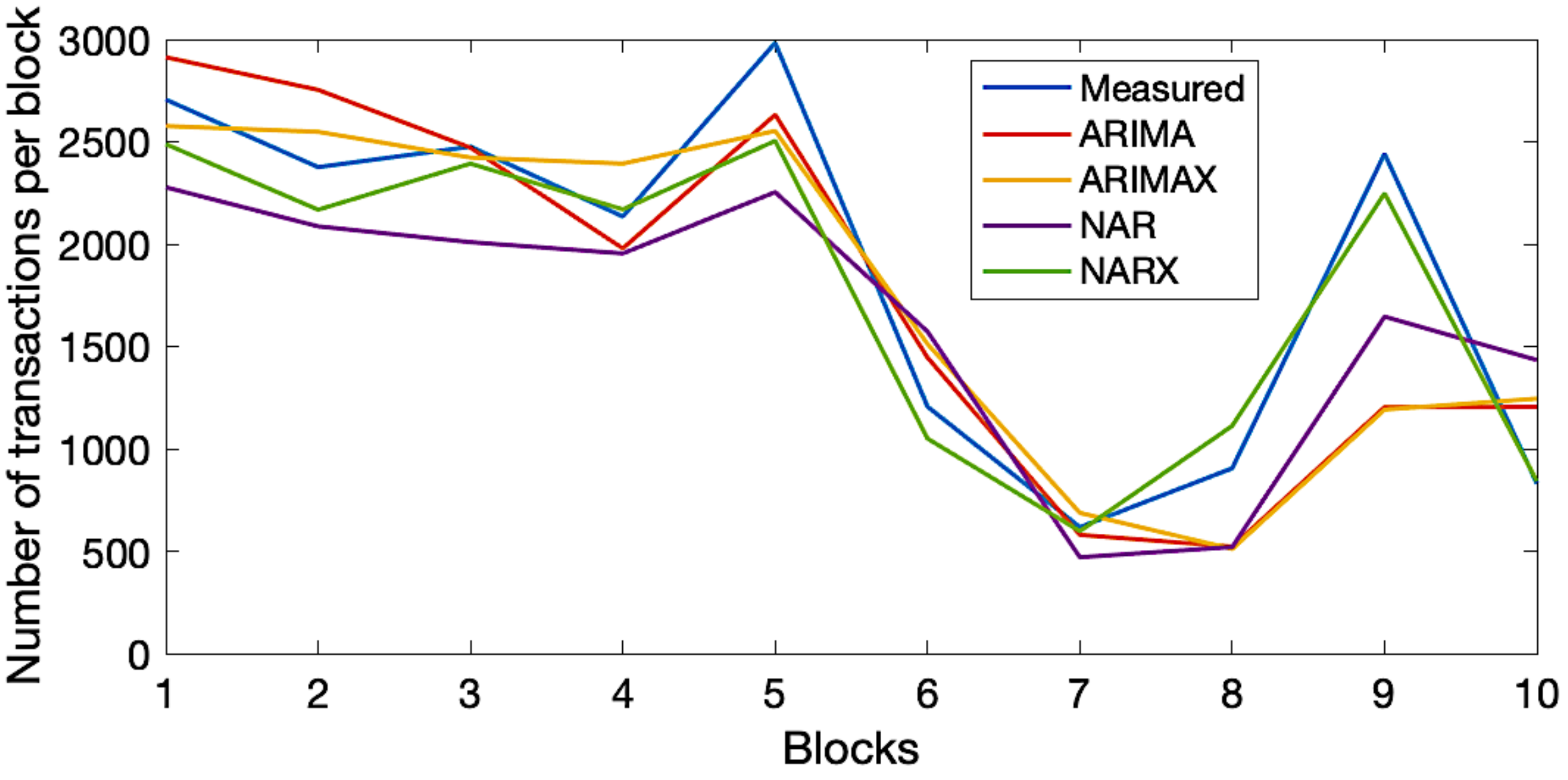}
   \label{weekendnut}
   
    }
    \label{weekend}
  \caption{Sample prediction results}  
\end{figure}

Table \ref{sourcecomparison} compares the performance of the various models in predicting the target block attributes: size $s_i$ and number $n_i$, where as a benchmark, the basic autoregressive (AR) model is also included.  For these models, the symbol $p$ is order of the autoregressive part, $d$ is the number of nonseasonal differences needed for stationarity, and $q$ is order of the moving average part. In this investigation, the values for $p=2$ and $q=2$ are calculated from autocorrelation and partial-autocorrelation plot, and we set $d=0$. MAE and RMSE are used to compare models' performance.
%Table \ref{sourcecomparison} reports the proposed models predicting the target blocks from the three cases.  In this table, the row represents the linear and nonlinear models, and the column represents three cases where the models' prediction performance in terms of the block size($s_i$) and the number of transactions inside a block ($n_i$). The symbol $p$ is order of the autoregressive part, $d$ is the number of nonseasonal differences needed for stationarity, and $q$ is order of the moving average part. The values for $p=2$ and $q=2$ are calculated from autocorrelation and partial-autocorrelation plot, we set $d=0$ is degree of first differencing involved. The MAE and MSE are used to compare models performance.
In addition, to give a more direct impression, we illustrate the prediction results by the models for randomly chosen ten consecutive weekend blocks, as an example, in Fig. \ref{weekendbsize} and Fig. \ref{weekendnut}. 

Table \ref{sourcecomparison}, Fig. \ref{weekendbsize} and Fig. \ref{weekendnut} indicate that, the prediction results by the considered forecasting approaches all follow the actual trend well. However, the models that additionally make use of the locally available information $x$, which are ARIMAX and NARX, generally produce better results than their counterpart models ARMA and NARX that do not have exogenous input. In addition, the AI-based models perform better than the classical autoregressive models under the same condition. Overall, NARX' performance is best, which is an encouraging finding for applying AI-based approaches in predicting the basic block attributes' values.   

{\bf Remark:} The alert reader may have noticed that among the three basic block attributes investigated in the exploratory study, we have left the fee $f_i$ out in the predictability study. This is simply because a large related literature exists, which will be discussed in the related work section, and the results therein show that the price can be excellently predicted. 

\begin{table*}[ht]
\caption{Forecasting performance of basic block attributes}  %%Linear and nonlinear models forecasting Performance from considering three cases while using the dataset as indicated by Table \ref{database}
%YM comment: Maybe also add fee.
\centering 
%\tiny
  \begin{tabular}{|l||l|l|l|l|l|l|l|l|l|l|}
    \hline
    \multirow{2}{*}{Models} &
      \multicolumn{3}{c|}{MAE} &
       \multicolumn{3}{c|}{RMSE}\\
      %\multicolumn{3}{c|}{RMSE} \\
      %\multicolumn{2}{c|}{C} \\
     & Weekend($s_i$,$n_i$) & Working($s_i$,$n_i$) & All($s_i$,$n_i$)  & Weekend($s_i$,$n_i$) & Working($s_i$,$n_i$) & All($s_i$,$n_i$) \\
    \hline
   
    AR(p)    & {0.53, 264} & {0.6, 117.35} & {0.5, 127.12} & {0.5, 122.14}& {0.5, 141.91} & {0.3, 264} \\ \hline
% ARMA(p,q)  & {0.056, 125.73} &  {0.04, 113.88} & {0.05, 118.53} & {0.01, 112.75} & {0.003, 103.05} & {0.004, 104.08} \\   \hline
%     ARMAX(p,q)    & {0.08, 127.12} &  {0.034, 125.86} & {0.045, 131.00} & {0.009, 111.63} & {0.002, 106.18} & {0.004, 110.52} \\    \hline
    ARIMA(p,d,q)    & {0.15, 15.373} &  {0.077,
12.840} & {0.13, 12.969} & {0.04, 12.461} & {0.01, 10.833} & {0.025, 10.942}\\
    \hline
      ARIMAX(p,d,q)    & {0.12, 13.364} &  {0.07, 12.092} & {0.06, 11.735} & {0.02, 11.052} & {0.006, 10.408} & {0.006, 10.408} \\
    \hline
%     MA(q)    & {0.29, 155.27} &  {0.11, 129.69} & {0.9, 132.31} & {0.11, 123.37} & {0.2,  243.51} & {0.13, 245.96} \\    \hline
     NAR(p)    & {0.01,
14.770} &  {0.06, 12.969} & {0.06, 12.840} & {0.03,
12.214} & {0.008, 11.275} & {0.008, 10.942} \\
    \hline
     NARX(p)    & {0.011,
10.942}&  {0.06, 10.471} & {0.013, 10.460} & {0.01,
10.121} & {0.006, 10.035} & {0.0003, 10.030} \\
    \hline
  \end{tabular}
 \label{sourcecomparison}
\end{table*}

\subsection{Block generation and transaction confirmation time}%{Block generation time and intensity}

\begin{figure}[th!]
    \centering
    \subfigure[Block generation time]
   {
    \includegraphics[width=0.45\linewidth, height=0.35\linewidth]{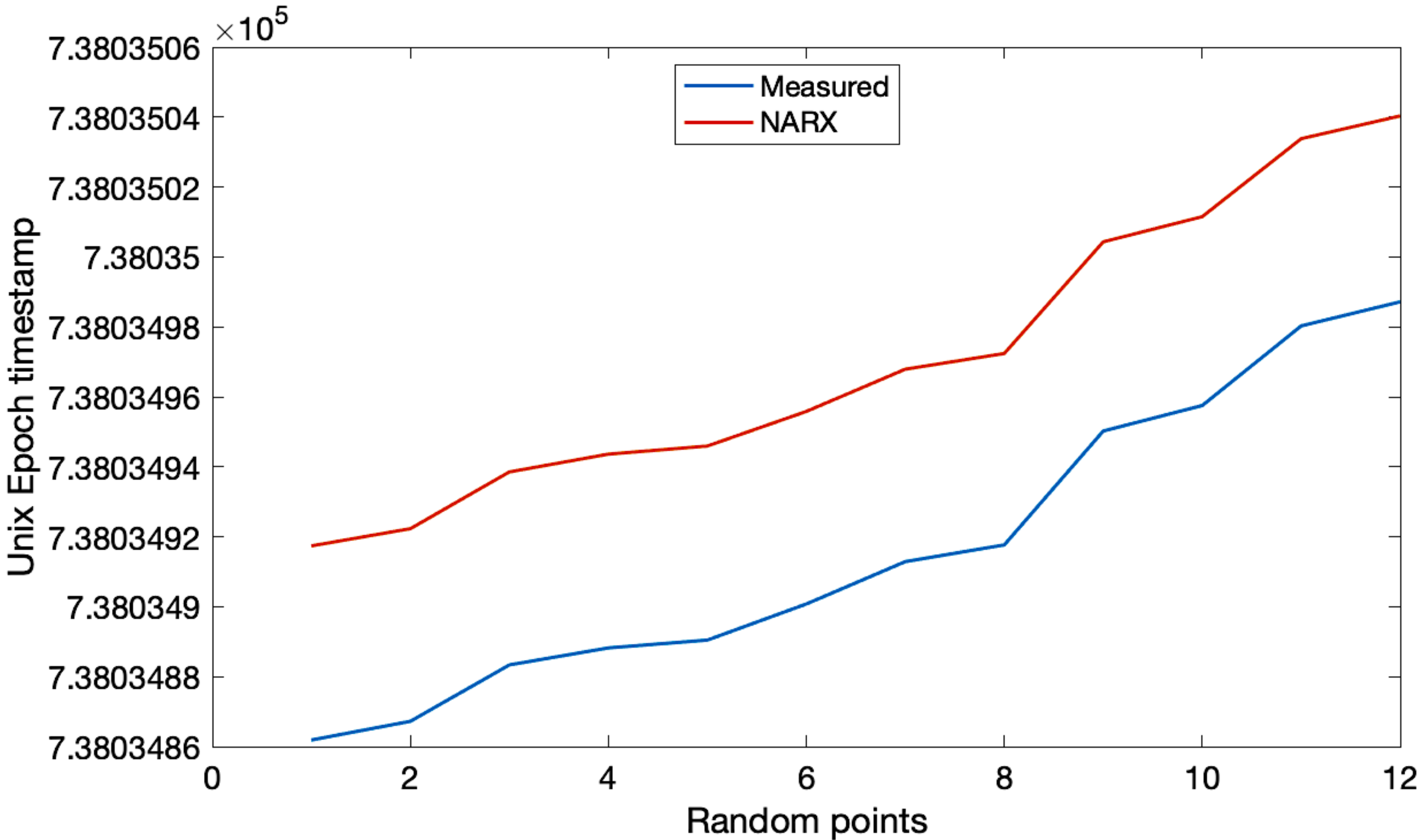}
   \label{performanceTime}
   }
   \subfigure[Transaction confirmation time]
    {
    \includegraphics[width=0.45\linewidth,height=0.35\linewidth]{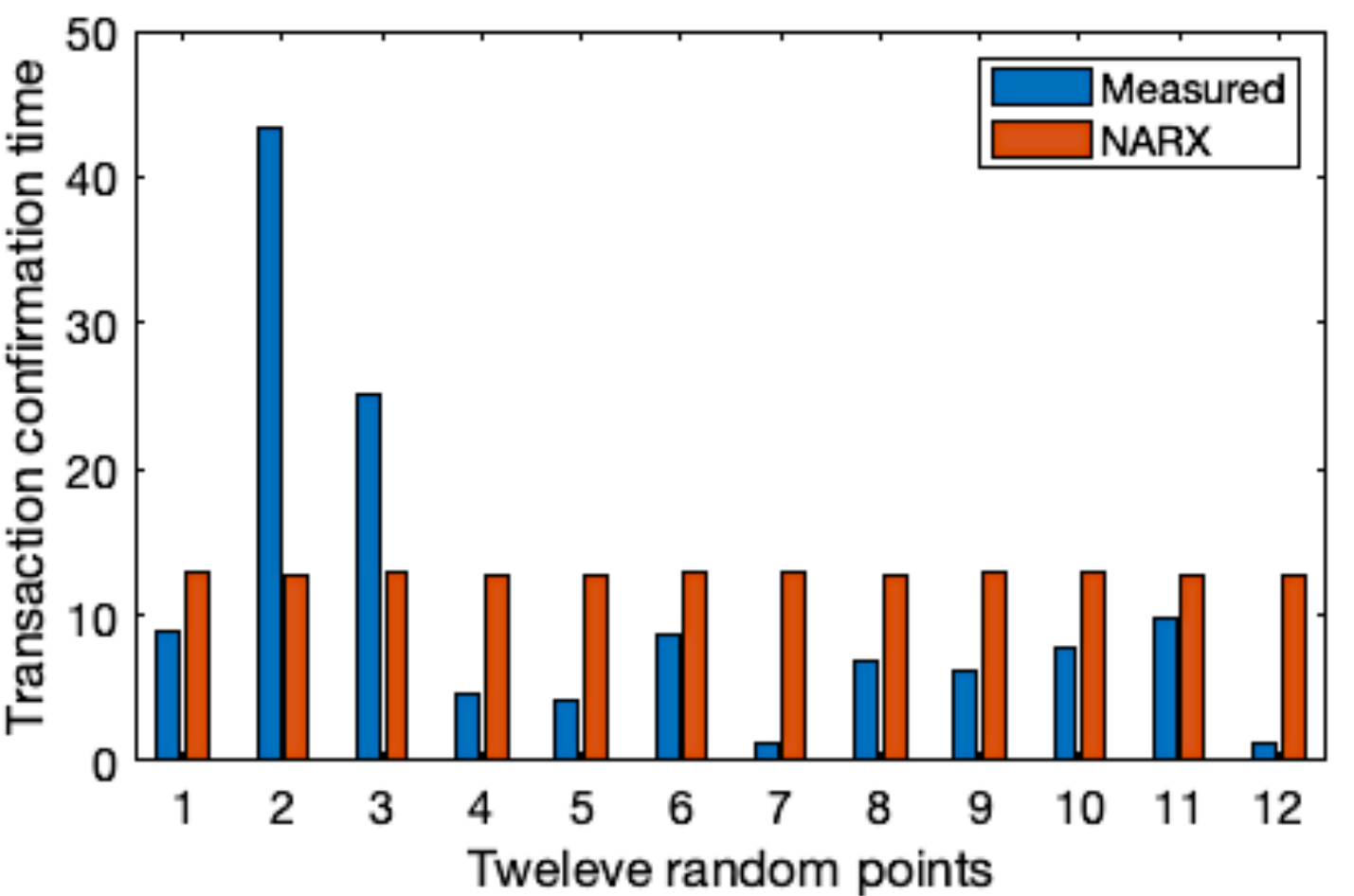}
   \label{conf}
  }
   % \label{kstest}
 \caption{Block generation and transaction confirmation time sample prediction}  
\end{figure}

Encouraged by the prediction results for the basic block attributes, we used the NARX model to test if block generation and transaction confirmation time can also be predicted. For predicting block generation, we used $T_i$ as the input while x=  $\{f_i, n_i, s_i, ms_i\}$ as the external input. Fig. \ref{performanceTime} reports the model's performance. For predicting transaction confirmation, we used transaction confirmation times as the input, while the size of the transactions and the fee associated are used as an external input.  Fig. \ref{conf} exemplifies the model's performance at a number of random points. 

As indicated by Fig. \ref{performanceTime} and Fig. \ref{conf}, the predication does not work. While this observation seems to be contradictory to the observation in predicting $s_i$ and $n_i$, a closer look at the characteristics of  block generation time and transaction confirmation time enables to explain. Reported in the exploratory analysis in Section \ref{sec-sa}, both the inter-block generation time and the transaction confirmation time has or can be closely approximated by an exponential distribution. Then, because of the memoryless property of exponential distribution, the likelihood of something happening in the future has little relation to whether it has happened in the past. Implied by this and as also confirmed by Fig. \ref{performanceTime} and Fig. \ref{conf}, any effort of predicting these two transaction handling aspects may, ``surpringly'', lead to no solid conclusion.

\begin{figure}[th!]
    \centering
    \subfigure[Block generation intensity with fixed time slot of 100 minutes]
   {
    \includegraphics[width=0.45\linewidth, height=0.37\linewidth]{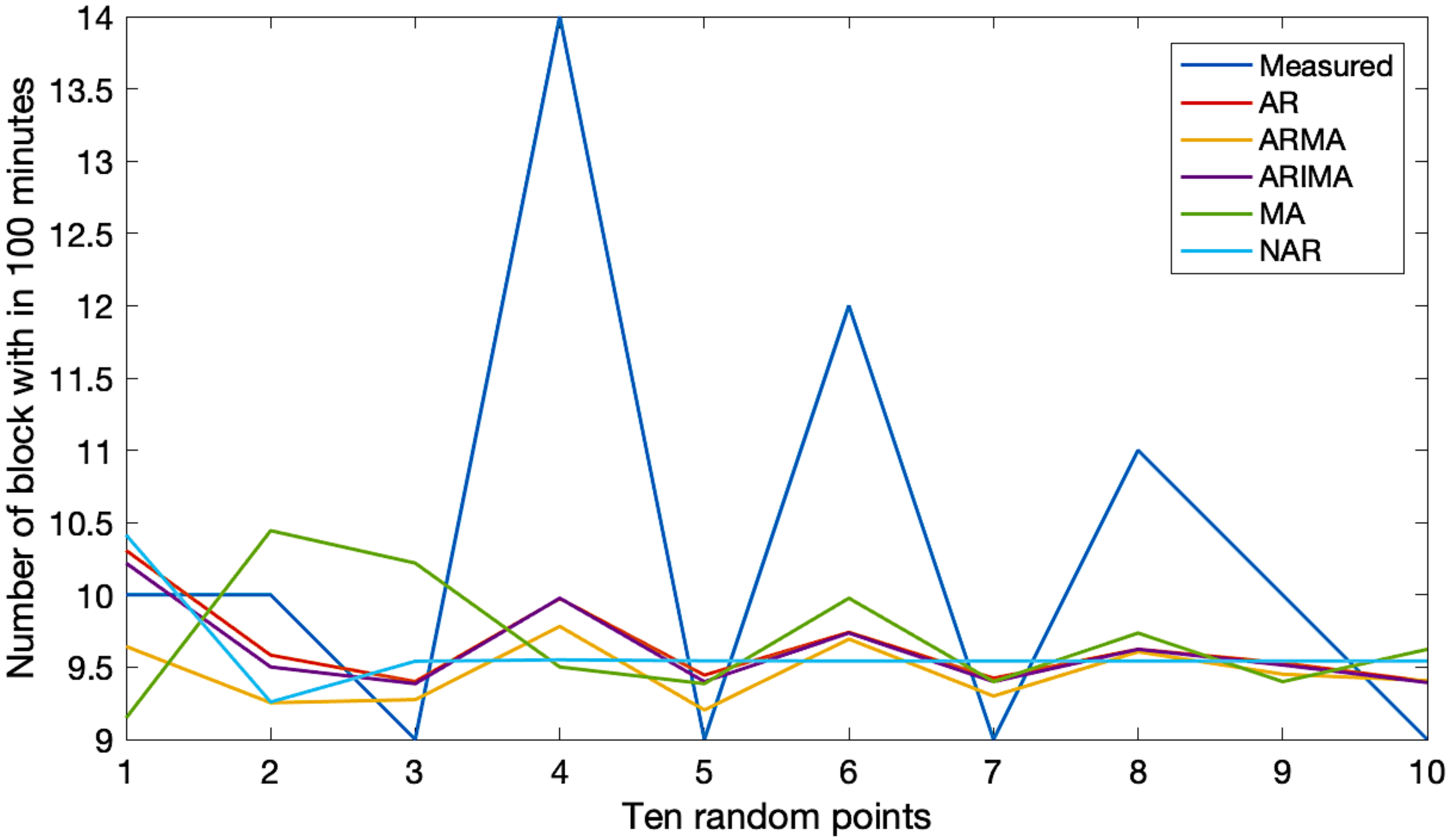}
   \label{per100}
   }
   \subfigure[Block generation intensity with fixed time slot of 1000 minutes]
    {
    \includegraphics[width=0.45\linewidth,height=0.37\linewidth]{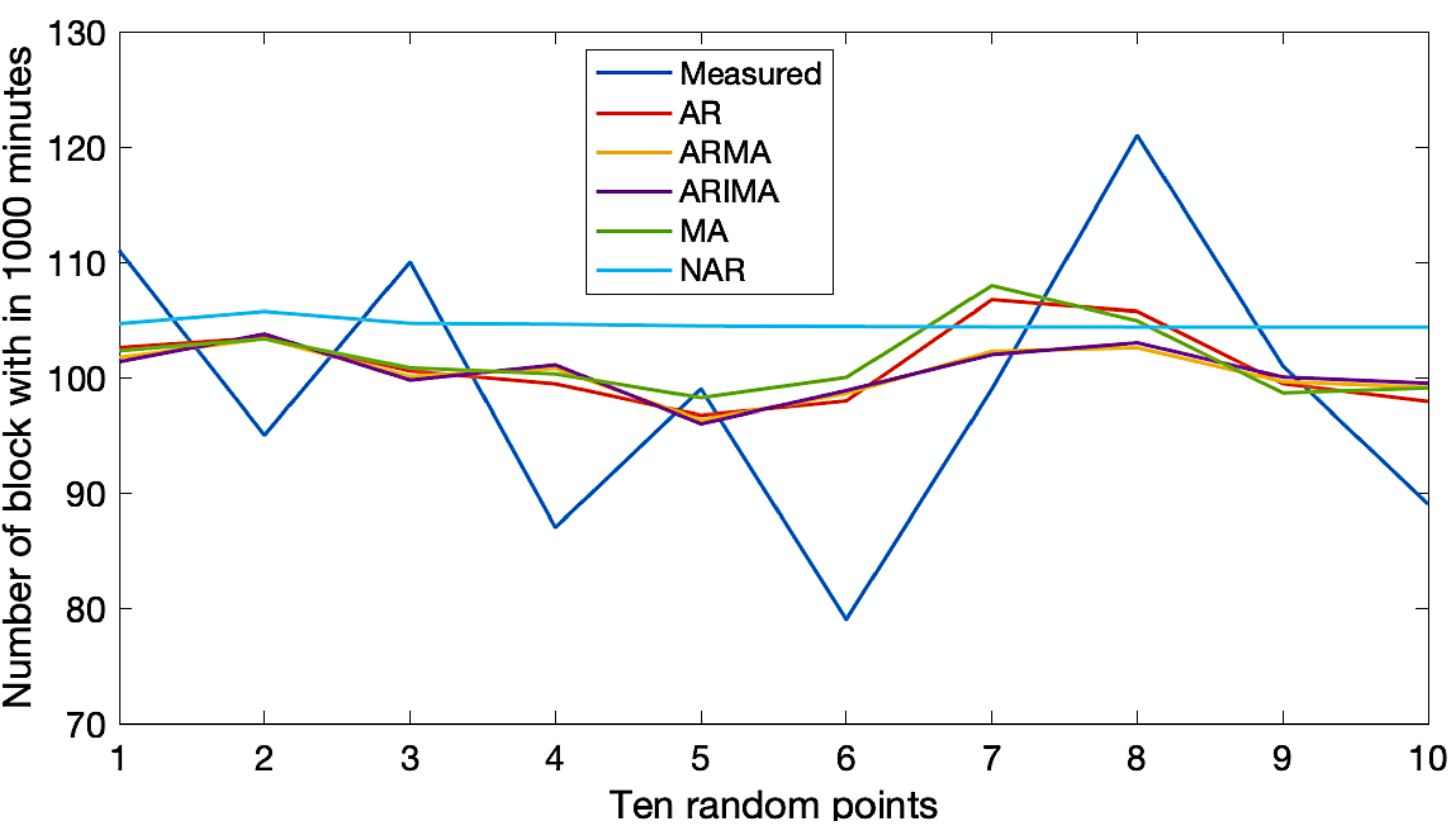}
   \label{per1000}
  }
   % \label{kstest}
 \caption{Block generation intensity sample prediction}  
\end{figure}

We conduct further investigations on predicting block generation intensity.  In this case, for the AI-based models, we only used NAR because we do not have additional input for NARX. To be in line with the counterpart exploratory investigation, we fixed the slot size of 100 and 1000 minutes and predicted the number of blocks within the slot, respectively.  Fig. \ref{per100} and \ref{per100} report the performance of both the classical autoregressive models and the AI-based NAR model. In general, the AR models follow the trend better than the NAR model. Nevertheless, all models struggle to perform better than the average. This, we believe, attributes largely from that while not exactly, the number of blocks in a time period can is approximately Poisson-distributed, as reported in Section \ref{sec-sa}. 

\subsection{Miner classification}
\begin{figure*}[th!]
   % \centering
    \subfigure[Weekend days]
    {
    \includegraphics[width=0.32\linewidth, height=0.35\linewidth]{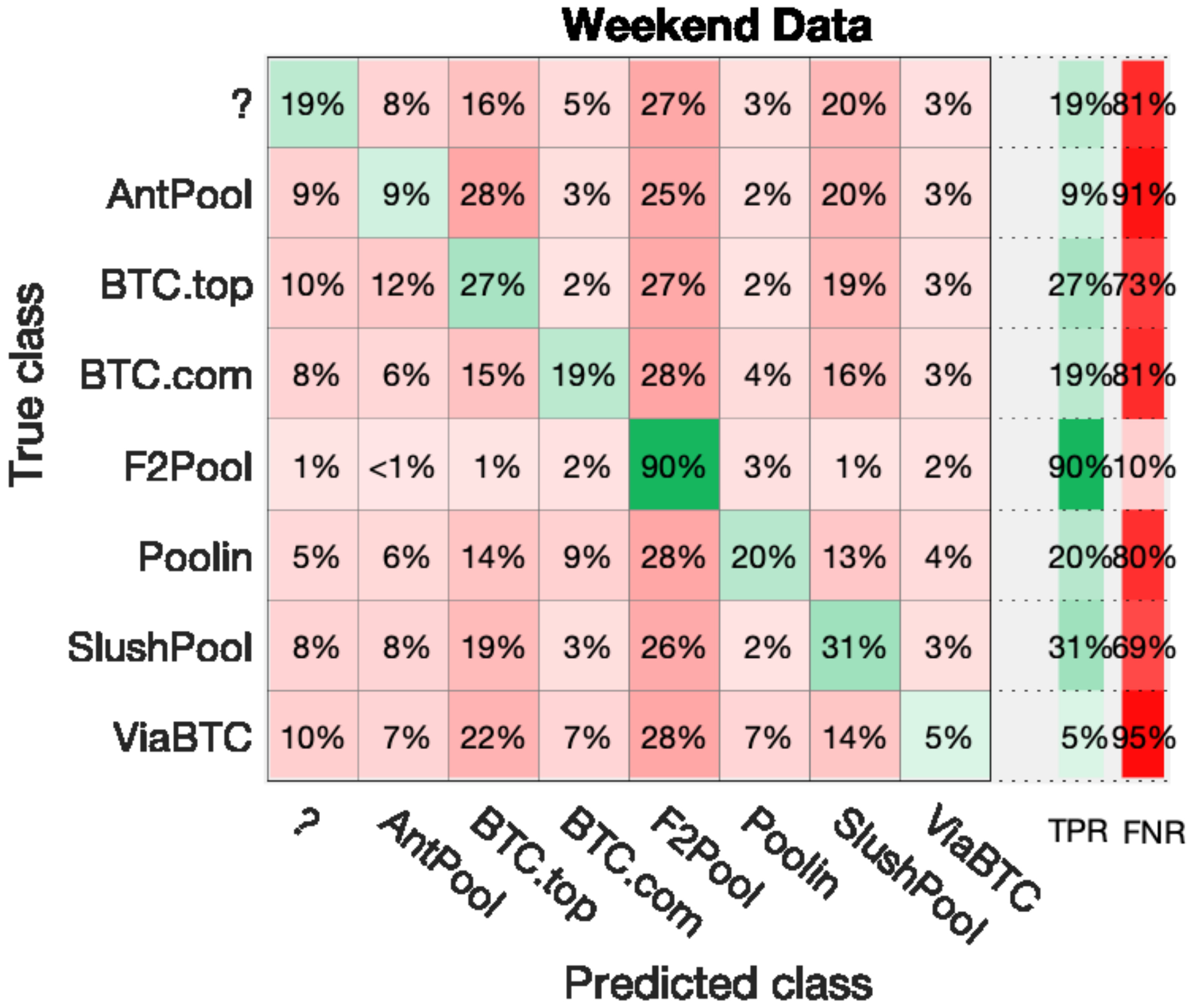}
   \label{Weekendminer}
    }
    \subfigure[Working days]
    {
    \includegraphics[width=0.32\linewidth, height=0.35\linewidth]{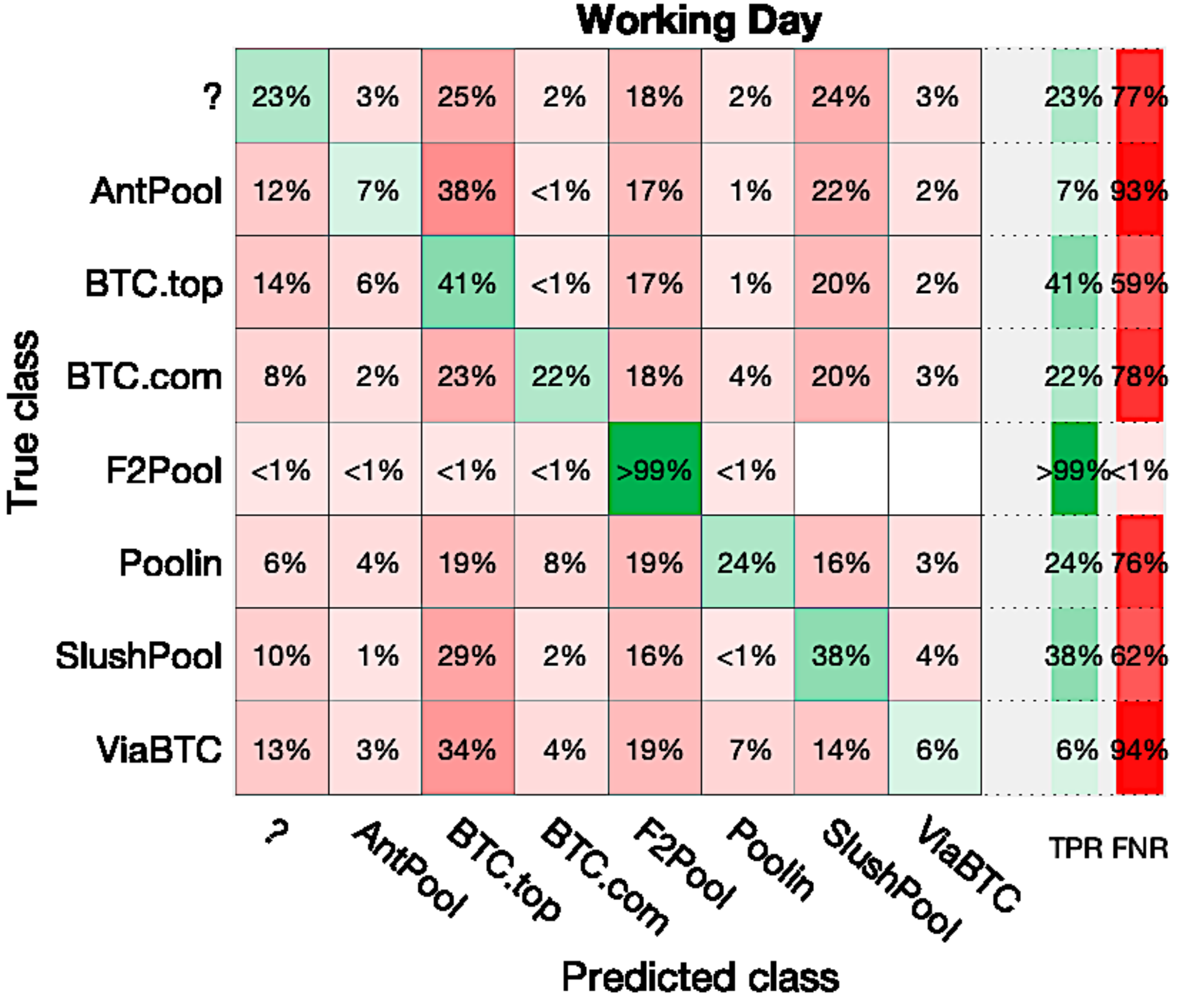}
   \label{Workingminer}
    }
    \subfigure[All]
    {
    \includegraphics[width=0.32\linewidth,height=0.35\linewidth]{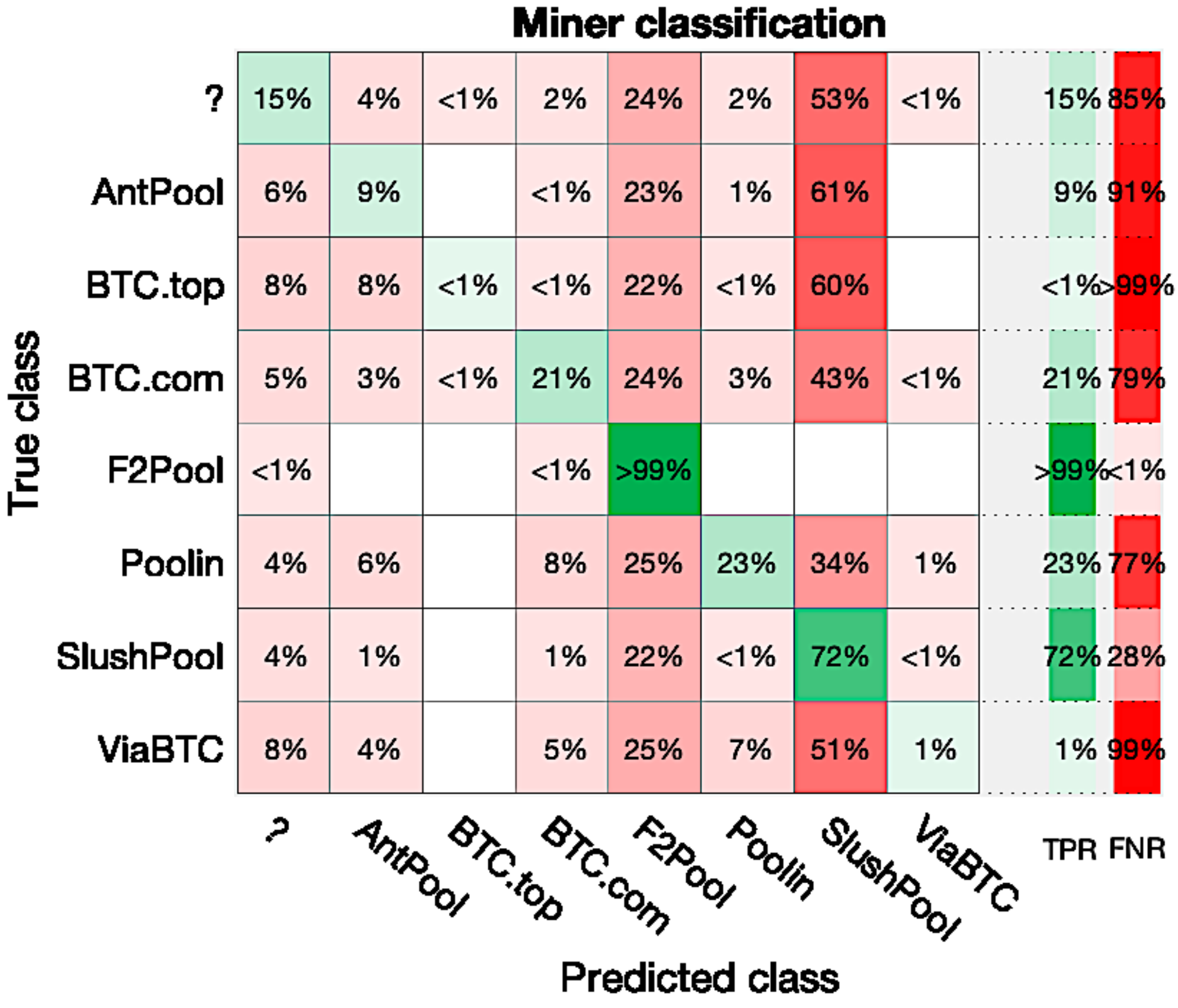}
   \label{confAll}
    }
%    \label{kstest}
  \caption{Confusion Matrix of major miners (RSUBoosted decision tree)}  
\end{figure*}
%The $f_i$, $s_i$, $n_i$, $ms_i$, and $Td_i$ were used to perform a throughput predictions.  However, the fact miners play a significant role in the throughput of the overall system. It triggers a question in which if these variables also determine/ classifies the block information's belongingness to the specific miners.  To answer such a question, we performed a decision-based classification of the major miners. The top five major miners are reported in Fig. \ref{MinerContr}.  It shows three of them have almost a close percentage of block contribution in working and weekend days except the Unknown(?) having more valid blocks.  

As we saw in the previous sections, the $f_i$, $s_i$, $Td_i$, $n_i$, and $ms_i$ have a significant effect on the evolution of the Bitcoin ledger.  Due to this, we use these feature sets to test if they can help infer a miner's relationship, and if some mining pools use some specified strategies while generating a block. To study these, we take two cases, first working and weekend days, and in the second case, considering all the data together.  The feature set, including $f_i$, $s_i$, $n_i$, and $Td_i$, is used to perform classifications of mining pools ($c_i$). As a remark, we have  also tried other features in the mempool state $ms_i$  but observed that they do not bring significant increase over the accuracy.

\subsubsection{Case-I (Working and Weekend day)}

The top-eight mining pools are used to detect the block generation behavior.  Fig.~ \ref{Weekendminer}  and Fig.~\ref{Workingminer} report that the major mining pools have a true positive rate (TP) more significant than the rest of the pools.  As Fig. \ref{Weekendminer}  and Fig. \ref{Workingminer}  report, the better model, the RSUBoosted decision tree with the booted method, shows a promising result classifying the F2Pool in better approximation relative to the other pools. As we can see from Fig. \ref{Weekendminer}  and Fig. \ref{Workingminer}, the TP for BTC.com, AntPool, and Poolin is smaller than 25\%, but for the SlushPool and BTC.TOP, it is more significant than 25\%.  Especially in the case of the public mining pool, Poolin, the false-negative rate is five times higher than the TP. This indicates the Poolin has less detectable block generation strategy than the rest. However, for SlushPool, it is has a block generation behaviour more distinguishable than the top five major mining pools.

\subsubsection{Case-II (All Data)}
\begin{figure}[ht!]
\centering
  \includegraphics[width=0.6\linewidth,height=0.33\linewidth]{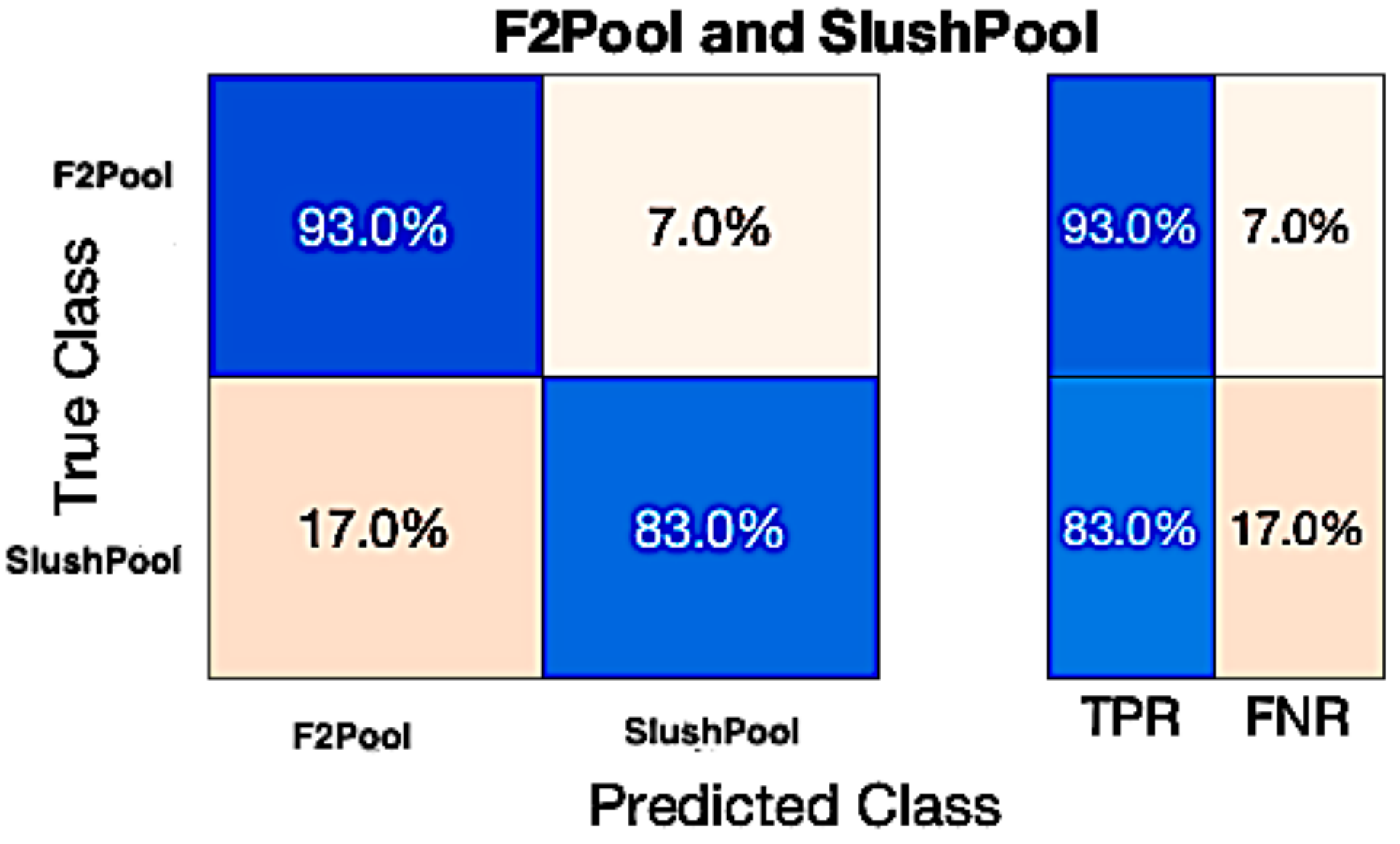}
  \caption{F2Pool and SlushPool}
  \label{two}
\end{figure}

\begin{figure}[th!]
    \centering
    \subfigure[F2Pool AUC curve]
    {
    \includegraphics[width=0.45\linewidth, height=0.43\linewidth]{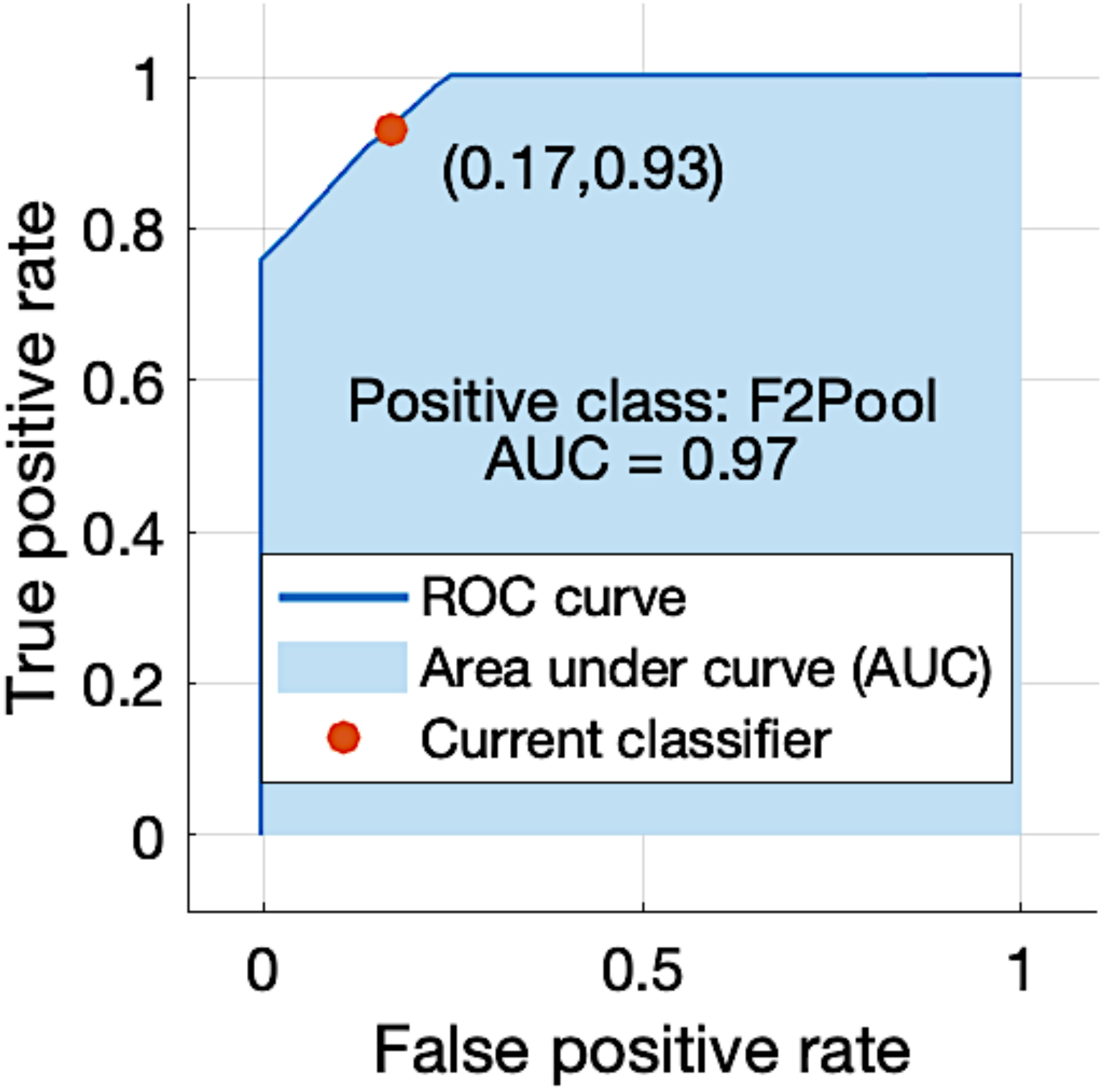}
   \label{auc}
    }
    \subfigure[? AUC curve]
    {
    \includegraphics[width=0.45\linewidth,height=0.43\linewidth]{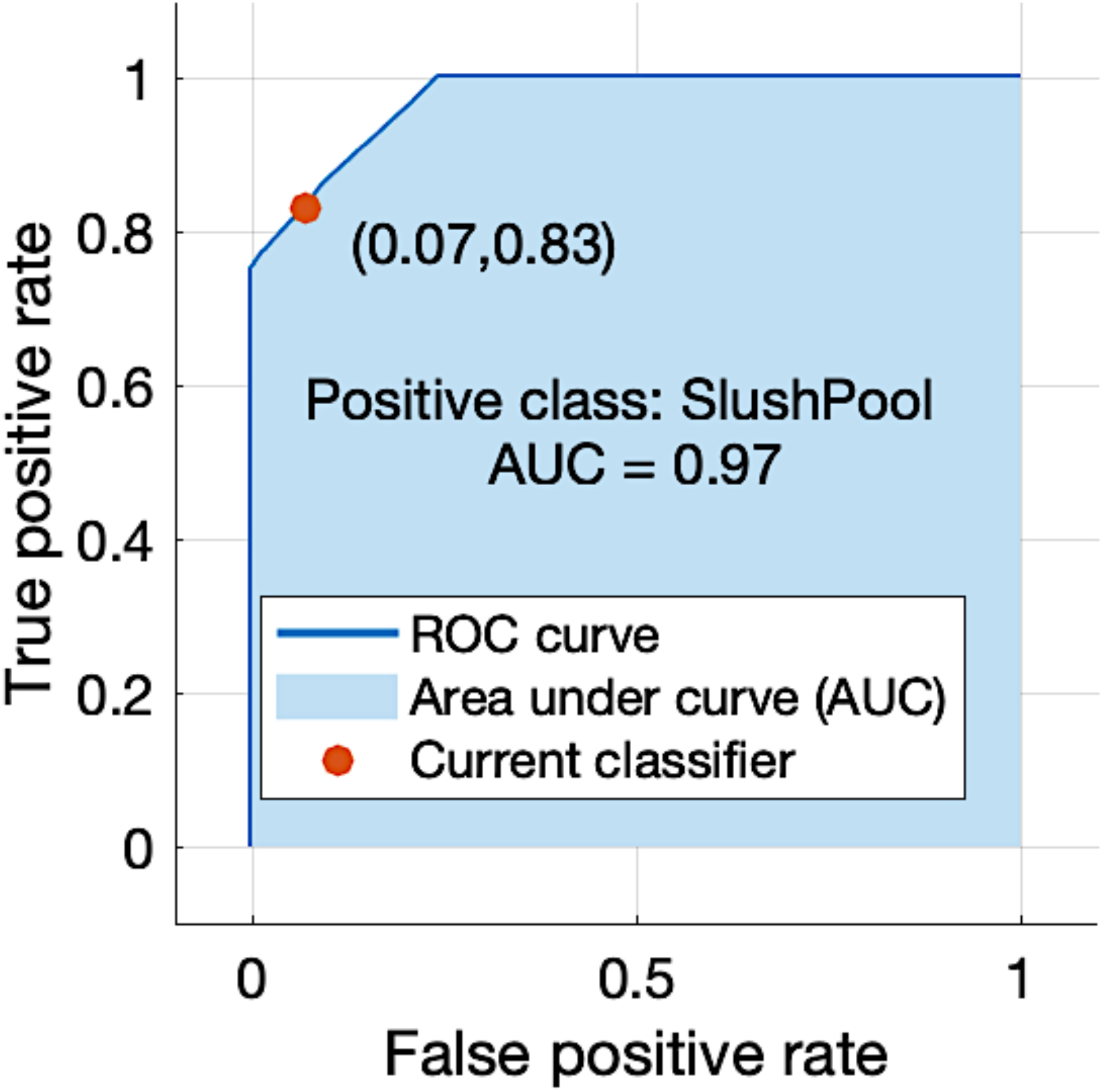}
   \label{auc2}
    }
   % \label{kstest}
  \caption{AUC curves for F2Pool and SlushPool}  
\end{figure}

The previous case showed that F2Pool was approximated very reasonably from the major mining pools. Fig. \ref{Weekendminer} and Fig.  \ref{Workingminer} report a confusion matrix illustrating the F2Pool and SlushPool having a higher positive rate than the rest of the mining pools. Additionally, Fig.\ref{confAll} reports that the two major mining pools, SlushPool and F2Pool, the TP are more significant than 70\%, which is 40\% more accurate than the first case for SlushPool. Similarly, the false-negative rate is less than 20\%, especially in F2Pool, which is even less than 3\%. 
To have a better understanding, we performed further investigation on only these two mining pools, F2Pool and SlushPool. The results are reported in Table \ref{twominer}, Fig. \ref{auc} and \ref{auc2}, and Fig. \ref{two}. Table \ref{twominer} compares the performance of the two DT methods. Due to better accuracy of the RSUBoosted-tree, it is used in Fig. \ref{auc} and \ref{auc2}, and Fig. \ref{two}. Specifically, the true-positive rate (TPR) and the false-negative rate (FNR) are shown in Fig. \ref{two}, and Fig. \ref{auc} and \ref{auc2} further illustrate the model accuracy in terms of AUC and ROC. %This also means these mining pools may use some strategies to select a transaction from the mempool than following the default fee per byte as Poolin. 

\begin{table}[ht!]
\centering
%\tiny
 \caption{Performance of classification between F2Pool and SlushPool} %Decision-tree models classification performance (F2Pool, SlushPool)
\begin{tabular}{|p{20mm}|p{10mm}|p{10mm}|p{12mm}|}
\hline
Models &  Accuracy   & Sensitivity & Miss rate\\
\hline
RSUBoosted-tree &  0.90   & 0.885 & 0.115\\
\hline
%Medium-tree &  0.884   & 0.882 &0.118\\ \hline
Boosted-tree & 0.883   & 0.881  &0.119\\
 \hline
\end{tabular}
\label{twominer}
\end{table}

\subsubsection{Discussion}
Fig. \ref{Weekendminer}, Fig. \ref{Workingminer}, and Fig. \ref{confAll} essentially show that other than for a few mining pools, particularly F2Pool, mining pools have a minimal positive classification rate, implying they are hard to distinguish. This is in line with Fig. \ref{InterBlock} in the exploratory analysis part, which shows that while the block generation distributions of other miners are similar, for F2Pool it is visually distinguishable from the others. We believe this characteristic difference has been explored by the decision tree approach in the classification. % Furthermore, Fig. \ref{Weekendminer}, \ref{Workingminer}, \ref{confAll} illustrates that the four high contributing mining pools having a positive classification rate greater than 10\%. 
In addition, a closer investigation as illustrated by Fig. \ref{confAll} and Fig. \ref{two} implies that the two major private mining pools P2Pool and SlushPool use different strategies that have caused their block generations with special properties making the classification with higher accuracy.  %Hence, other than the two private mining pools,  pools were difficult for the classifier to find a pattern that provides some insight into their block generation's strategies.

\section{Related Work} \label{sec-stateArt}
\subsubsection{Statistical analysis of transaction handling characteristics} 
While a lot of such analysis results are available, e.g., various Bitcoin statistics~\cite{Btc}, block propagation delay~\cite{infoProp2013}, block arrival process \cite{blkArrival2020}, transaction rate and confirmation time~\cite{DiscreteBlockchain}~\cite{ trasactionConfirmation}, %most are directly generated or derived from the information carried on the Bitcoin blockchain.
we focus on fundamental aspects underlying transaction handling and particularly their distributions, different from the literature. Through analyzing these distributions, we have been able to reason some seemly surprising observations in the predictability study. In addition, very few results in the literature take into account information that is only locally available. In this sense, the work \cite{TransBitcoin} is most related. However, except for inter-block generation time fitting, which is similar as we already highlighted, the other results are not found in \cite{TransBitcoin}, due to different focuses of \cite{TransBitcoin} and the present work.

\subsubsection{Forecasting transaction handling characteristics}%Throughput prediction
%Several works employed data collection to perform analysis. For instance, 
%Artificial intelligence methods have been implemented to study the pattern of some feature sets that show nonlinear relationships while having some hidden pattern, where 
The focus of the literature has been on bitcoin price. For instance,  Huisu Jang and Jaewook Lee \cite{Bayesian} developed a neural network-based forecast model on the volatility of a Bitcoin price and extended analysis to identify the best feature set that gives more information about the Bitcoin price  process.  Similarly, Edwin Sin and Lipo Wang \cite{bitcoinPrice} implemented an artificial neural network to predict the next Bitcoin price and the amount of profit that could be gained by making such predictions. Shah et al. \cite{bayesianBit} considered the Bayesian regression method to predict the price of Bitcoin.  Pavel Ciaian et al. \cite{economics} estimates Bitcoin price formation based on a linear model by introducing several factors such as market forces, attractiveness for investors, and global macro-financial factors. Greaves et al. \cite{greaves} analyzed the Bitcoin blockchain data to predict the price of Bitcoin using SVM and ANN, which score 55\% accuracy. Similarly,  models such as Random Forest, SVM, and Binomial Logistic algorithms are used to predict short-term Bitcoin price and achieve a high accuracy result of 97\% in \cite{madan}. To the best of our knowledge, no previous work combines the feature sets to predict the transaction handling characteristics focused in this paper.

%Most of the available literature focuses on analyzing price volatility to determine the fluctuations over time, such as \cite{DeepLN}, \cite{BitNN}, \cite{Bayesian}, \cite{Deeplearning}, \cite{Four}, \cite{prediBit}, \cite{VolatalityBitcoin}, \cite{bitcoinPrice}, \cite{BitLSTM}.  Motivated by such works, we use supervised machine learning techniques (ML) to analyze different measurements’  contribution. To the best of our knowledge, no work combines the feature sets to study the Bitcoin next block characteristics.

\subsubsection{Mining pool classification}
%Solving the mathematical puzzle brings a reasonable reward; however, it is also difficult for a single miner to find a nouce with expectation only every few days or even every few weeks \cite{incentive}.  Miners join mining pools to increase the gain and the chance of finding nonce through cooperation computation effort.  To increase revenue, miners can perform block withholding attacks on the mining pool, or mining pools can also use miners to infiltrate into other mining pools and conduct block withholding attacks on other mining pools \cite{bitcoinGame}.  
There have been some research works that studied block withholding and unfair distribution of reward.  For instance, Schrijvers et al. \cite{incentive} analyzed the incentive compatibility of the Bitcoin reward mechanism. In their model, a miner can decide between honest mining and delaying her found blocks' submission. They proved that the proportional mining reward mechanism is not incentive compatible.  Eyal \cite{eyaldilemma} computed the pools' optimal strategy in the block withholding attack and their corresponding revenues. It was demonstrated that the no-pool-attack strategy is not a Nash equilibrium in these games because if none of the pools run the attack, one pool can increase its revenue by launching the attack.  
Luu et al. \cite{luu} experimentally demonstrate that block withholding can increase the attacker's revenue. They do not address the question of mutual attacks. Courtois and Bahack \cite{courtois} have recently noted that a pool can increase its overall revenue with block withholding if honest pools perform all other mining. We consider the general case where not all mining is performed through public pools and analyze situations where pools can attack one another. M. Salimitari et al. \cite{Profit} used prospect theory to predict a miner's Profit from joining one of the major mining pools. The hash rate power, total number of the pool members, reward distribution policy of the pool, electricity fee in the new miner's region, pool fee, and the current Bitcoin value are used to predict which pools are profitable specific miners.

Most mining pool studies do either emphasis on {\bf (i)} block withholding \cite{bitcoinGame} \cite{MininGamemodel} or  {\bf (ii)} unfair distribution of rewards \cite{socialMining} \cite{miningEvolu} \cite{Intstrategy} \cite{bitcoinGameCorr}, but none or little has been investigated to detect the major mining pools with hidden block generation strategies.  Our work tries to further investigate these block formation strategies, by introducing decision tree to distinguish one of the major mining pools following having a detectable block formation strategy.

\section{Conclusion}\label{sec-con}

An exploratory analysis on fundamental  transaction handling characteristics of Bitcoin is conducted, together with a novel analysis on their predictability. The results from the former have been used to help reason the findings from the latter. Specifically, the focused block attributes include the size, the number of transactions and the fee. In addition, block generation and transaction confirmation, two fundamental processes resulted from transaction handling, are investigated. Furthermore, the contribution of miners to these attributes and processes is particularly taken into consideration.  

The results show that while it is possible to use measurement-based collected data in predicting the basic attributes of the next block with reasonable accuracy, care is needed in predicting block generation and transaction confirmation. While the latter seems contradicting the expectation from the former, the explanation is supported and implied by results from the exploratory analysis. Additionally, it shows that combining internal and external factors enables better performance in prediction / classification. Furthermore, although it is difficult to distinguish among mining pools through prediction in general, the investigation shows that F2Pool is well distinguished from the others. A closer investigation in the exploratory analysis shows that block generation of F2Pool has a distribution with visible characteristic difference, implying that it has used a different strategy than the other miners.   
These results shed new light and may also be considered by users and miners when deciding their transaction strategies.

%We employ existing linear and nonlinear models to forecast consecutive bitcoin blocks and classify the significant mining pools.  This approach's main originality is learning a representation of bitcoin blocks and the models' ability to extract the relationship between feature sets with mining pools while considering internal and external factors.  The prediction's effectiveness is tested by dividing the dataset into three categories. At the same time,  the division is implying the difference in throughput in all the cases. Similarly, the classification effectiveness in finding a hidden pattern by the mining pools in block generation is also tested by considering two cases. The first case considers weekend and working day, and in the second case, considering all the data.

%Our analysis shows that it is possible to use measurement-based collected data in predicting the characteristics of the next block with reasonable accuracy and distinguish some major private mining pools' behavior.  Additionally, it shows that combining internal and external factors enables a much better prediction/classification. Such results suggest that the forecasting throughput in bitcoin and analyzing the miners may significantly benefit from proposed approaches.

\AtNextBibliography{\footnotesize}
{\footnotesize \printbibliography}

\end{document}